\documentclass[usegraphicx,usenatbib]{mn2e}

 at 10truept

\title [The impact of halo shapes on the bispectrum in cosmology]
{\vglue-3.0truecm \vglue 2.5truecm The impact of halo
shapes on the bispectrum in cosmology}

\author
[{\it Smith, Watts \& Sheth}]
{\parbox[t]{\textwidth}{
    R.~E.~Smith$^{1}$ \thanks{res@astro.upenn.edu}
    P.~I.~R.~Watts$^{2}$ \thanks{pwatts@astro.uni-bonn.de}
    \& R.~K.~Sheth$^{1}$\\
      {\small
      $^1$ Department of Physics and Astronomy,
      University of Pennsylyvania,
      209 South 33rd Street, Philadelphia, PA 19104, USA. \vspace{-0.2cm} \\
      $^2$ Institut f. Astrophysik u. Extraterrestrische Forschung,
      Universit\"at Bonn, Auf dem H\"ugel 71, 53121 Bonn, Germany.
      \vspace{-0.2cm}\\
    }
  }
}

\def\bk{{\bf k}}
\def\br{{\bf r}}
\def\bx{{\bf x}}

\def\dirac{{\delta^D}}

\def\abc2{{\left(\frac{ab}{c^2}\right)}}

\def\nbarg{\bar{n}_{\rm g}}
\def\a{{\bf a}}
\def\d3x{{d^3\!x}}

\def\be{\begin{equation}}
\def\ee{\end{equation}}
\def\rhob{\bar{\rho}}

\def\nbar{\bar{n}}
\def\simless{\mathbin{\lower 3pt\hbox
  {$\rlap{\raise 5pt\hbox{$\char'074$}}\mathchar"7218$}}}   
\def\simgreat{\mathbin{\lower 3pt\hbox
   {$\rlap{\raise 5pt\hbox{$\char'076$}}\mathchar"7218$}}}  
\def \scs{\scriptscriptstyle}
\def \L{\,{\scs{\rm L}}}

\def\ba{\begin{eqnarray}}
\def\ea{\end{eqnarray}}

\def\bess0{{\rm J_0}}
\def\sbess0{{\rm j_0}}
\def\cyc{{\rm cyc}}
\def\1H{1{\rm H}}
\def\2H{2{\rm H}}
\def\3H{3{\rm H}}

\def\Nmax{N_M}
\def\rmi{\rm i}
\def\cent{\rm cent}
\def\sat{\rm sat}
\def\CW{\mathcal{W}}

\newcommand{\avec}{\mbox{\boldmath $\mathcal{E}$}}

\begin{document}

\maketitle


\begin{abstract}\vspace{-0.3cm}\\
We use the triaxial halo model formalism of \citet{SmithWatts2005} to
investigate the impact of dark matter halo shapes on the cosmological
bispectrum. Analytic expressions for the dark matter distribution are
derived and subsequently evaluated through numerical integration. Two
models for the ellipsoidal halo profiles are considered: a toy model
designed to isolate the effects of halo shape on the clustering alone;
and the more realistic model of Jing \& Suto (2002).  For equilateral
$k$-space triangles, we show that the predictions of the triaxial
model are suppressed, relative to the spherical model, by up to
$\sim7\%$ and $\sim4\%$ for the two profiles respectively.  When one
considers the reduced bispectrum as a function of triangle
configuration it is found to be highly sensitive to halo shapes on
small scales. The generic features of our predictions are that,
relative to the spherical halo model, the signal is suppressed for
$k$-vector configurations that are close to equilateral triangles and
boosted for configurations that are colinear.  This appears to be a
unique signature of halo triaxiality and potentially provides a means
for measuring halo shapes in forthcoming cosmic shear surveys.

The galaxy bispectrum is also explored. Two models for the halo
occupation distribution (HOD) are considered: the binomial
distribution of Scoccimarro et al. (2001) and the Poisson satellite
model of Kravtsov et al. (2004). Our predictions show that the galaxy
bispectrum is also sensitive to halo shapes, although relative to the
mass the effects are reduced.  The HOD of Kravtsov et al. is found to
be more sensitive. This owes to the fact that the first moment of the
occupation probability is a steeper function of mass in this model,
and hence the high mass (more triaxial) haloes are more strongly
weighted. Interestingly, the functional form of the configuration
dependent bispectrum is, modulo an amplitude shift, not strongly
sensitive to the exact form of the HOD, but is mainly determined by
the halo shape. However, a combination of measurements made on
different scales and for different $k$-space triangle configurations
is sensitive to both halo shape and the HOD.

\end{abstract}



\section{Introduction}

In our current picture of structure formation, small Cold Dark Matter
(CDM) density fluctuations, seeded during an early epoch of cosmic
inflation, collapse through gravitational instability. From an initial
spectrum of fluctuations that is almost scale-invariant, the collapse
proceeds in a hierarchical way with fluctuations on small scales
collapsing at early times to form dense clumps of dark matter
(haloes), and with those on larger scales collapsing at later
times. This gives rise to a web-like network of filaments and clusters
-- the \emph{`cosmic web'} \citep{BondMyers1996}.  Current
observational constraints strongly suggest that the growth of
structure takes place in a Universe that is flat and which at late
times is dominated by an unknown Dark Energy component that drives an
accelerated Universal expansion \citep{Spergeletal2003}. Building a
complete statistical description of the clustering pattern of the dark
matter is consequently of great importance: this may provide detailed
information concerning the physics of the dark matter, the statistics
of the primordial fluctuations and also valuable insight into the
nature of dark energy.  Direct comparison of the dark matter
distribution with that of the galaxies reveals the salient physics of
galaxy formation.

However, the dynamics of structure formation are nonlinear and
analytic solutions for the equations of motion are only tractable for
highly idealized cases.  Robust calculation of the CDM density field
therefore requires recourse to numerical integration via $N$-body
methods. Over the past decades significant effort has been invested in
performing and analyzing large numerical simulations.  The results of
this have led to a number of semi-empirical formulae that describe the
dark matter halo phenomenology to high accuracy. In particular, halo
abundances \citep{ShethTormen1999,Jenkinsetal2001}, large-scale bias
\citep{MoWhite1996,ShethTormen1999}, density structures
\citep{NavarroFrenkWhite1997,Bullocketal2001,Poweretal2003,JingSuto2002},
internal velocity distributions and large scale velocity bias
\citep{Shethetal2001a,Shethetal2001b,ShethDiaferio2001} are now well
characterized. More recently, attention has been focused on
quantifying halo shapes (JS02), substructure phenomenology
\citep{Mooreetal1999,Klypinetal1999,Ghignaetal2000} and effects on the
halo due to baryon physics
\citep{Gnedinetal2004,Kazantzidisetal2004}. Collectively, these ideas
have lead to the development and evolution of the halo model paradigm
for structure formation.

In the most basic formulation of the halo model
\citep{Seljak2000,PeacockSmith2000,MaFry2000}, one assumes that all of
the dark matter in the Universe is confined to virialized, spherical
haloes that possess some distribution in mass, some universal density
structure and a large-scale bias derivable from perturbation theory,
all of which are characterizable through the halo mass. The clustering
statistics over all scales then conveniently break down into terms
involving correlations between haloes and correlations within
haloes. This approach has had a high degree of success in modelling
the low-order spatial clustering statistics of the mass.  Most
advantageously, the halo model can also be used to make predictions
for the clustering statistics of the galaxy distribution, which in
general can be thought of as a biased sampling of the underlying mass
distribution.  Indeed, in the halo picture the concept of bias is
replaced with that of the Halo Occupation Distribution (HOD), which
dictates the number and position of galaxies within dark matter haloes
\citep{BerlindWeinberg2002}. Moreover, the HOD connects in a
satisfying manner to both the theoretical predictions of galaxy
formation models \citep[][and
others]{Bensonetal2000,Seljak2000,Scoccimarroetal2001,BerlindWeinberg2002}
and direct observational reality \citep[][and
others]{PeacockSmith2000,BerlindWeinberg2002,Yangetal2003}.

Whilst the agreement between the halo model and numerical simulations
is in general good at the two-point level, significant differences
have been reported at the three-point level
\citep[SSHJ;][]{Fosalbaetal2005}.  Although a recent study by
\citet{Wangetal2004} showed that the spherical halo model could be
modified to provide a better fit to simulation data, if a finite
volume correction and a somewhat arbitrary spatial exclusion scale for
the haloes were taken into account.  However, discrepancies still
remain discernible on both small and large scales. It is currently
believed that the various inconsistencies between the halo model and
N-body simulations are manifestations of the breakdown of key
assumptions in the model.  Thus, given the significant interest in
this halo based approach, it is of crucial importance to understand
the validity of all its approximations. In particular, these issues
must be addressed if we are to take seriously the halo model as a tool
for precision cosmology.

One possible erroneous assumption is that dark matter haloes are
spherical, or rather, that we may work with the spherical averages of
the density profiles. However, it has long been known that the haloes
found in numerical simulations are more closely described as triaxial
ellipsoids \citep{BarnesEfstathiou1987,Frenketal1988,Warrenetal1992}.
Recently, JS02 have shown, through high resolution $N$-body
simulations, that the inclusion of halo shape information into the
modelling of the density structure leads to improved fitting
functions. With these ideas in mind, some key questions can now be
asked: To what extent do the shapes of the dark matter haloes affect
the clustering statistics of the mass distribution? Is there any
observable affect on the galaxy clustering?

In a recent paper, \citet[][hereafter SW05]{SmithWatts2005} laid out
the foundations for performing halo model calculations with triaxial
haloes: the `triaxial halo model'. They then went on to calculate the
importance of halo shapes on the dark matter power spectrum, the
Fourier transform of the two-point correlation function. They reported
that the effect of halo triaxiality was to suppress the power spectrum
on small/nonlinear scales at the level of between $\sim$ 5\% and
$\sim$ 15\%, dependent on the precise choice of density profile model.

A number of other modifications to the halo model may lead to similar
effects on the power spectrum. Particular examples are the inclusion
of a stochastic concentration parameter in halo density profiles
\citep{CoorayHu2001} and the treatment of halo substructure
\citep{ShethJain2003,Dolneyetal2004}. More recently, some have
suggested that changes to the halo boundary definition may affect
significant changes to the clustering
\citep{TakadaJain2003a,Fosalbaetal2005}. Clearly, if one wishes to
disentangle these effects one must look to higher order statistics. In
this paper we apply the triaxial halo model formalism of SW05 to the
task of computing the bispectrum, the Fourier space analog of the
three-point correlation function. The bispectrum is considered to be the
lowest order spatial statistic that is sensitive to the shapes of
structures. We will demonstrate that this measure can indeed be used
to differentiate between changes to halo density profiles and changes
to halo shapes.

The paper breaks down as follows: In Section \ref{sec:theory} we
review some theoretical notions and lay down the bones of the triaxial
halo model formalism. In Section \ref{sec:3point} we provide the
details of our derivation of the three-point correlation and bispectrum
for the mass, presenting the results in Section \ref{sec:results}.  In
Section \ref{sec:galaxies} we switch from mass clustering to galaxy
clustering and repeat the analysis for galaxies.  Finally, we discuss
our findings and draw our conclusions in Section
\ref{sec:conclusions}.

Throughout we assume a flat, dark energy dominated cosmological model
with equation of state manifest as a cosmological constant. We take
$\Omega_m=0.3$ and $\Omega_{DE}=0.7$, where $\Omega_m$ and
$\Omega_{DE}$ are the ratios of the density in dark matter and dark
energy to the critical density, respectively. We use the linear power
spectrum of \citet{Efstathiouetal1992} with normalization
$\sigma_8=0.9$ and shape parameter $\Gamma\equiv\Omega_m h=0.21$, where
$h=0.7$ is the dimensionless Hubble parameter.


\section{Theoretical background}\label{sec:theory}

\subsection{three-point spatial clustering statistics}\label{ssec:3ptGen}

In this paper we are concerned with the three-point spatial statistics of
the density contrast field $\delta(\br)$. This field is defined
through the relation
\be\rho(\br) = \rhob\left[1+\delta(\br)\right],\label{deltadeff}\ee
where $\rho(\br)$ is the physical density and $\rhob$ is the density
of the background. The three-point correlation function, $\zeta$, is
defined as the ensemble average of $\delta$ measured at three points
in space. This can be written
\be
\zeta(\br_1,\br_2,\br_3) \equiv  \left<
\delta(\br_1)\,\delta(\br_2)\,\delta(\br_3)\right>,\ \label{xi3deff}
\ee
where the angled brackets denote the ensemble average.  For
homogeneous random fields $\zeta$ obeys translational invariance:
\be
\zeta(\br_1,\br_2,\br_3)=\zeta(\br_1+\br_0,\br_2+\br_0,\br_3+\br_0),
\label{eq:3ptHomog}\ee
where $\br_0$ is an arbitrary position vector. For isotropic random fields,
$\zeta$ also obeys rotational invariance:
\be \zeta(\br_1,\br_2,\br_3)=\zeta({\mathcal R}\br_1,{\mathcal
R}\br_2,{\mathcal R}\br_3),\label{eq:3ptIsotropy}\ee
where ${\mathcal R}$ represents an arbitrary coordinate rotation. For
universes that obey the cosmological principle these two properties
must hold. Lastly, $\zeta$ must also be invariant under parity
transformations:
\be
\zeta(\br_1,\br_2,\br_3)=\zeta(-\br_1,-\br_2,-\br_3)\ .
\label{eq:3ptParity}
\ee 

In what follows, we will not deal explicitly with $\zeta$, but instead
work with its Fourier transformed counterpart, the bispectrum $B$,
\be \left<\delta(\bk_1)\delta(\bk_2)\delta(\bk_3)\right> =
(2\pi)^3 B(\bk_1,\bk_2,\bk_3)\, \dirac(\bk_1+\bk_2+\bk_3)\ ,\ee
where $\delta(\bk)$ is the Fourier transform of $\delta(\br)$ and
$\dirac$ is the Dirac delta function. From the argument of the delta
function we obtain the triangle condition, i.e that the sum of the three
wave-vectors $\bk_1$, $\bk_2$ and $\bk_3$ form the null vector.  The
bispectrum and three-point correlation function are explicitly related
through
\[
\zeta(\br_1,\br_2,\br_3) = \int \frac{d\bk_1}{(2\pi)^{3}}
\frac{d\bk_2}{(2\pi)^{3}} \, \, B(\bk_1,\bk_2,-\bk_1-\bk_2)
\]
\be \hspace{1.6cm}
\times
\exp{[-\rmi \bk_1\cdot(\br_1-\br_3)]} \exp{[-\rmi \bk_2\cdot(\br_2-\br_3)]}
\label{CF2Bisp}
.\ee
From the reality of $\zeta$, we find that
\be B(\bk_1,\bk_2,\bk_3)=B^{*}(-\bk_1,-\bk_2,-\bk_3) \ ,
\ \label{eq:biprop1}\ee
where $^*$ denotes complex conjugation and henceforth the triangle
condition is to be assumed.  If the $\delta$-field is statistically
homogeneous and isotropic, then following equation
(\ref{eq:3ptParity}) the bispectrum also obeys parity invariance:
\be B(\bk_1,\bk_2,\bk_3)=B(-\bk_1,-\bk_2,-\bk_3)\ .\ee
This, combined with the reality constraint of equation
(\ref{eq:biprop1}) implies that the bispectrum is real. Lastly,
following equation (\ref{eq:3ptIsotropy}) the bispectrum is also
rotationally invariant:
\be B(\bk_1,\bk_2,\bk_3)=B({\mathcal R}\bk_1,{\mathcal R}\bk_2,{\mathcal R}
\bk_3) \ . \label{eq:biprop2}\ee
%


\subsection{Ellipsoidal systems}\label{ssec:ellipscoords}

In what follows we will be concerned with the clustering properties of
triaxial dark matter haloes. We therefore define some important
relations for these objects \citep[see][for a full
treatise]{Chandrasekhar1969}.  Consider a heterogeneous triaxial
ellipsoid that has semi-axis lengths $a$, $b$ and $c$, where $a \leq b
\leq c$, and orthogonal principle axis vectors
$(\hat{\bf{e}}_a,\hat{\bf{e}}_b,\hat{\bf{e}}_c)$.  We define the
triaxial coordinate system $(R,\Theta,\Phi)$ with respect to the
principle axes of the halo, where $\hat{\bf{e}}_c$ is taken to be in
the $z$-direction of a standard Cartesian system.  The radial
parameter $R$ traces out thin iso-density shells, or homoeoids, and
the parameters $\Theta$ and $\Phi$ are the polar and azimuthal angles
respectively. In this system the Cartesian components can be written
\be x = \frac{a}{c} R \cos \Phi \sin \Theta \ ; \ y = \frac{b}{c} R
\sin \Phi \sin \Theta \ ; \ z = R \cos\Theta.
\label{ellipdef}\ee
It is to be noted that the ellipsoidal angles differ from those of the
spherical coordinate system. The parameter $R$ can be related to the
Cartesian coordinates and axis ratios through
\be \frac{R^2}{c^2} = \frac{x^2}{a^2} + \frac{y^2}{b^2} +
\frac{z^2}{c^2} \label{eq:isodef}\ .\ee
The benefits of this choice of coordinate system are now apparent: if
all homoeoidal shells are concentric, and if one picks coordinates
with the same axis ratios as the triaxial ellipsoid, then the density
run of the ellipsoid can be described by a single parameter:
\be \rho(\br)\rightarrow\rho(R)\ .\label{eq:rexchange}\ee
Furthermore, the mass enclosed within some iso-density cut-off scale
$R_{\rm cut}$, can be obtained most simply by
\be M=\int_{R_{\rm cut}(\br)} d\br \rho(\br)=4\pi \frac{ab}{c^2}
\int_0^{R_{\rm cut}} dR R^2 \rho(R) \ ,\ee
where the ellipsoidal coordinates have allowed us to circumvent the
problem of evaluating complicated halo boundaries.

It is now also convenient to define what we mean by a halo: any object
that has a volume averaged over-density 200 times the background
density is considered to be a gravitationally bound halo of dark
matter. This leads directly to the following relation between the
mass, radius and axis ratios
\be M_{200}=\frac{4}{3}\pi R_{200}^3 \abc2 200 \rhob \
.\label{eq:M200}\ee
The above definition was adopted in order to be consistent with the
mass-function of \citet{ShethTormen1999}.


\subsection{Triaxial halo model}\label{sec:formalism}

We next summarize the theoretical ingredients of the triaxial halo
model presented in SW05. Consider a density field comprised
entirely of ellipsoidal dark matter haloes of different masses.
The density run of each halo is specified by some universal
profile, the exact details of which are not important at this
stage save that the general properties discussed in the previous
section are satisfied. Each halo may therefore be characterized by
a set of stochastic variables. These are the position vector for
the halo centre of mass, $\bx$, the mass $M$, the principle axis
lengths $(a,b,c)$, and the direction vectors for these axes
$(\hat{\bf{e}}_a, \hat{\bf{e}}_b, \hat{\bf{e}}_c)$. We express
these last two sets of variables compactly by using the notation
$\a\equiv(a,b,c)$ and $\avec\equiv(\hat{\bf{e}}_a,\hat{\bf{e}}_b,
\hat{\bf{e}}_c)$. Thus, the density at any point $\br$ can be
expressed as a sum over the $N$ haloes that comprise the field
\be \rho(\br) = \sum_i^N M_i\, U(\br-\bx_i , M_i , \avec_i , \a_i),
\label{sumoverhaloes}
\ee
where $U$ is the mass normalized density profile and where sub-scripts
$i$ denote the characteristics of the $i^{\rm th}$ halo.

The correlation functions of the density field follow directly from
taking ensemble averages of products of the density at different
points in space. To compute the ensemble averages we integrate over
the joint probability density function (hereafter PDF) for the $N$
haloes that form the density field, and sum over the probabilities for
obtaining the $N$ haloes \citep[see][for a similar
approach]{McClellandSilk1977} %
\[ \langle \cdot\cdot\cdot \rangle \equiv \sum_{j} p(N_j|V) \int
\prod_{i=1}^{N_j} dM_i\;d\bx_i\;d{\bf a}_i\;d\avec_i\]
\be \hspace{0.3cm} \times \
p(M_1,..,M_{N_j},{\bf x}_1,..,{\bf x}_{N_j},{\bf a}_1,..,{\bf
  a}_{N_j}, \avec_1,..,\avec_{N_j}|N_j)\ . \label{eq:ensemble}\ee
Provided the volume of space considered is large, then $p(N|V)$ is
very sharply spiked around $N=\nbar V\gg 1$, where $\nbar$ is the mean
number density of haloes. We restrict our study to this case
only.

The integrals over $\avec$ in equation (\ref{eq:ensemble}) represent
averages over all possible orientations of the halo. The orientation
of the halo frame can be specified relative to a fixed Cartesian basis
set through the Euler angles.  These represent successive rotations of
the halo frame about the $z$--axis by $\alpha$, the $y'$--axis by
$\beta$, and the $z''$--axis by $\gamma$
\citep[see][]{MathewsWalker1970}.  Hence, the process of averaging
over all halo orientations can be performed by integrating over all
Euler angles. We thus make the following transformation
$\avec\rightarrow (\alpha,\beta,\gamma)$.

Finally, we will require explicit expressions for the joint PDF for a
single halo's characteristics, $p(\bx,M,\avec,\a)$. If we assume that
a halo's orientation, position and mass are independent random
variables, and that the halo axis ratios are dependent on mass only
(see e.g JS02), then we may write
\be  p(\bx,M,\avec,\a) =
\frac{1}{V}\frac{n(M)}{\nbar}p(\avec)p(\a|M) , \label{probone} \ee
where $n(M)$ is the mass function of haloes. If the orientation of a
halo is uniformly random on the sphere then $p(\avec) = 1/8\pi^2$ (see
SW05 for further details).


\section{Clustering of triaxial haloes}\label{sec:3point}

\subsection{The three-point correlation function}

We are now in a position to compute the three-point correlation function
$\zeta(\br_1,\br_2,\br_3)$ in the triaxial halo model. This is done by
applying the machinery of equations (\ref{sumoverhaloes}) and
(\ref{eq:ensemble}), to the definitions given by equation
(\ref{xi3deff}) in conjunction with equation (\ref{deltadeff}) -- see
also SW05. We find that, as for the case of the standard halo model,
$\zeta$ can be expressed as the sum of three terms
\be \zeta=\zeta^{\1H}+\zeta^{\2H}+\zeta^{\3H} \
,\label{eq:xi3halomodel}\ee
where we have suppressed the position vector dependence for
simplicity. Here $\zeta^{\1H}$, $\zeta^{\2H}$ and $\zeta^{\3H}$ give
the contributions for the case where: all three points are within the
same halo; two points are within the same halo and the third is within
a separate halo; all three points are within distinct
haloes. Throughout, we refer to these contributions as the 1-, 2- and
3-Halo terms. These have the following forms:
\[
\zeta^{\1H}= \frac{1}{\rhob^3 8\pi^2} \int dM\;
d \bx\;d{\bf a}\; d\avec\; \;M^3 \; n(M) p({\bf a}|M)\] \be
\hspace{1cm}\times\prod_{i=1}^{3}\left\{ U(M,\bx-\br_i,{\bf
a},\avec)\right\}\;
\label{xi3-1H}\ ;
\ee
\[\zeta^{\2H}= \frac{1}{\rhob^3(8\pi^2)^2}\int
\prod_{i=1}^{2} \left\{ dM_i\, d \bx_i\, d{\bf a}_i\, d\avec_i\,
 n(M_i) \,p({\bf a}_i|M_i)\right. \]
\[ \hspace{1cm}\times \left. \  U(\bx_i-\br_i,M_i,{\bf
a}_i,\avec_i) \right\} M_1^2 \,M_2 \]
\be \hspace{1cm}\times \ \,
U(\bx_1-\br_3,M_1,{\bf a}_1,\avec_1)\; \xi^s(1,2)+\cyc\ ;\label{xi3-2H}\ee
\[ \zeta^{\3H} =\frac{1}{\rhob^3(8\pi^2)^3}\int \prod_{i=1}^{3}
\left\{ dM_i\; d \bx_i\; d{\bf a}_i\;d\avec_i \; M_i n(M_i)\, p({\bf
a}_i|M_i) \right. \]
\be \hspace{1cm} \times \, \left. U({\bf
x}_i-\br_i,M_i,{\bf a}_i,\avec_i)\right\} \,\zeta^s(1,2,3)\ \label{xi3-3H}.\ee
In deriving this result we have used the fact that the joint PDF for
the characteristics of three haloes may be rewritten
\[ p(1,2,3) =  p(1)\,p(2)\,p(3)\left[1+\xi^s(1,2)+\xi^s(2,3)
\right.\]
\be \hspace{1.5cm} \left.+\xi^s(3,1)+\zeta^s(1,2,3)\right]\ ,\ee
where we have used the short-hand notation $p(1)\equiv
p(\bx_1,M_1,\a_1,\avec_1)$ etc. The quantity $p(1)$ is the 1-Halo PDF,
as given by equation (\ref{probone}), while $\xi^s(1,2)$ and
$\zeta^s(1,2,3)$ are the two- and three-point \emph{seed} correlation
functions for triaxial haloes.

In the work of SW05 it was shown that the effects of halo alignments
on the mass clustering statistics are insignificant, even for the
highly symmetrized case where all of the haloes are perfectly aligned.
We therefore take the orientation and shape of each halo to be
statistically independent of each and every other halo. Under this
condition the joint halo PDF for the three-point halo properties reduces
to
\[
p(1,2,3) = p(1)\,p(2)\,p(3)\left[1+\xi^s(M_1,M_2,\bx_1,\bx_2)\right.
\]
\[
\hspace{1.5cm}+\xi^s(M_2,M_3,\bx_2,\bx_3)+\xi^s(M_3,M_1\bx_3,\bx_1)
\]
\be
\left. \hspace{1.5cm}+\zeta^s(M_1,M_2,M_3,\bx_1,\bx_2,\bx_3)\right]\
,\ee
where the seed correlation functions can now be specified in the usual
way \citep{MoWhite1996,ShethTormen1999}.


\subsection{The Bispectrum}\label{ssec:bispectrum}

The bispectrum is most directly obtained through Fourier transforming
equation (\ref{eq:xi3halomodel}). From the linearity of $\zeta$, we
see that $B$ is also composed of three terms
\be B_{123}=B^{\1H}_{123}+B^{\2H}_{123}+B^{\3H}_{123}\ ,\ee
where $B^{\1H}_{123}$ is the contribution from the 1-Halo term etc.,
and where we have adopted the short-hand notation $B_{123}\equiv
B(\bk_1,\bk_2,\bk_3)$. In the subsequent sub-sections we derive
explicit relations for these quantities.


\subsubsection{The 1-Halo term}

Fourier transforming equation (\ref{xi3-1H}) leads us to the 1-Halo
contribution to the bispectrum. This is compactly written
\be  B^{1{\rm H}}_{123}=\frac{1}{\rhob^3}
\int dM\, M^3\,n(M) \,W_{123}(M)\ , \label{eq:bi1H}\ee
where we have introduced the (three-point) window function
\be W_{123}(M)\equiv \frac{1}{8\pi^2}\!\int d\a \, d\avec \,p(\a|M)
\!\!\prod_{i=\{1,2,3\}}\!\! U(\bk_i,M,\a,\avec),\label{eq:w123def}\ee
where $U(\bk_i,M,\a,\avec)$ is the Fourier transform of the mass
normalized triaxial halo profile. In its present form equation
(\ref{eq:w123def}) is of limited practical use, since its computation
requires one to solve a very high dimensional numerical integral. In
Section \ref{ssec:orientation} we show how this expression can be
greatly simplified.


\subsubsection{The 2-Halo term}

The Fourier transform of equation (\ref{xi3-2H}) gives the 2-Halo
contribution to the bispectrum. This is succinctly written
\[ B^{2H}_{123} = \frac{P_{\L}(k_2)}{\rhob^3} \int \prod_{i=\{1,2\}}
\left\{dM_i\, n(M_i)\,b(M_i) \right\}\] \be \hspace{1cm}\times M_1^2\,
M_2\,W_{13}(M_1) \, W_{2}(M_2) + \cyc\ ,\label{eq:bi2H} \ee
where $W_i$ and $W_{ij}$ are the 1- and two-point window functions
defined by the relations
\be W_{i}(M)\equiv\frac{1}{8\pi^2}\int d\a p(\a|M)\int d\avec
U(\bk_i,M,\a,\avec)\ ;\label{eq:w1def}\ee
and
\be W_{ij}(M) \equiv \frac{1}{8\pi^2}\int d\a \, d\avec \,p(\a|M)
\prod_{\nu=\{i,j\}} U(\bk_{\nu},M,\a,\avec) \ ;\label{eq:w12def}\ee
and where we have written the Fourier transform of the halo seed
two-point correlation function as $P^s(k,M_1,M_2)=b_1(M_1)
\,b_1(M_2)\,P_{\L}(k)$, where $P_{\L}$ is the linear power spectrum
and $b_1(M)$ is the first linear halo bias parameter
\citep{MoWhite1996,ShethTormen1999}. Again, these expressions will be
simplified following the discussion in Section \ref{ssec:orientation}.


\subsubsection{The 3-Halo term}

Finally, the 3-Halo contribution to the bispectrum is obtained through
Fourier transforming equation (\ref{xi3-3H}),
\[ B^{3H}_{123}=\frac{1}{\rhob^3}\int \prod_{i=1}^{3}
\left\{dM_i \,M_i\, n(M_i)\,W_i(M_i)\right\}\] \be \hspace{1cm}\times\
B^s_{123}(M_1,M_2,M_3) \ ,\label{eq:bi3H}\ \ee
where $W_{i}(M)$ is given by equation (\ref{eq:w1def}). The function
$B^s_{123}(M_1,M_2,M_3)$ is the bispectrum of halo seeds and is given
by (SSHJ)
\[ B^s_{123}(M_1,M_2,M_3)=b_1(M_1)\;b_1(M_2)\;b_1(M_3)\]
\be \hspace{1cm}\times\left[B^{\rm PT}_{123}+
\left\{\frac{b_2(M_3)}{b_1(M_3)}
\,P_{\L}(k_1)\,P_{\L}(k_2)+\cyc\right\}\right]\ ,\ee
where $b_2(M)$ is the second order halo bias factor.  The quantity
$B^{\rm PT}_{123}$ is the second order Eulerian perturbation theory
bispectrum \citep{Fry1984,JainBertschinger1994},
\be B^{\rm PT}_{123}=2F_2(\bk_1,\bk_2)P_{\rm L}(k_1)\,P_{\rm L}(k_2)+
  \rm \; \cyc \ , \ee
and where
\be
F_2(\bk_1,\bk_2)=\frac{5}{7}+\frac{1}{2}\cos\theta_{12}(k_1/k_2+k_2/k_1)+
\frac{2}{7}\cos^2\theta_{12}\ ,\ee
with $\cos\theta_{12}=\bk_1\cdot\bk_2/k_1k_2\,$.


\subsection{Simplification of halo window functions}
\label{ssec:orientation}

Having written down the basic results for the bispectrum, we now
consider in more detail the expressions for the halo window functions
defined by equations (\ref{eq:w123def}), (\ref{eq:w1def}) and
(\ref{eq:w12def}). So far, these have been left in terms of
$U(\bk_1,M,\a,\avec)$, the Fourier transform of the triaxial
profile. As mentioned, numerical evaluation of these quantities in
their present form would be very cumbersome, owing to the dependence
of the halo profile on the orientation and shape of the halo. This
means that the (full 3-D) Fourier transforms cannot be pre-computed
but instead must be evaluated inside a 5-D integral.  The following
arguments serve to simplify these expressions considerably.

Consider first the one-point window function given by equation
(\ref{eq:w1def}), on inverse Fourier transforming the halo profile we
find,
\[ W_{i}(M)=\frac{1}{8\pi^2}
\int d\a p(\a|M) \int d\avec\, d\br\, U(\br,M,\a,\avec)\] \be
\hspace{1cm}\times\ \exp(-\rmi\bk_i\cdot\br) \ .\ee
If we now construct a Cartesian coordinate system about the direction
vectors of the semi-axes of the halo, then we may transform from that
system to a system of ellipsoidal coordinates where the isodensity
surfaces of the halo are given by equation (\ref{eq:isodef}). Thus, in
this new basis set the density profile simply becomes a function of
the ellipsoidal radial parameter $R$: $U(\br)\rightarrow U(R)$.

The orientation average can now be performed directly by rotating the
reference Cartesian coordinate system through all possible values of
the Euler angles. Hence, the components of the $k$-vector are modified
through each infinitesimal rotation $(\delta\alpha,\delta \cos
\beta,\delta\gamma)$, which leads us to
\[ W_{i}(M)=\frac{1}{8\pi^2}\int d\a p(\a|M)\frac{ab}{c^2} \int
dR\,R^2 U(R,M,\a) \] \be
\hspace{1cm}\times \int d\alpha\, d(\cos\beta)\, d\gamma \int
d\hat{\bf R} \exp\left[-\rmi{\bk' _i\cdot\br}({\bf R})\right]
\label{window} \ ,\ee
where $\bk'={\mathcal R}(\alpha,\beta,\gamma)\bk$ and ${\mathcal
R}(\alpha,\beta,\gamma)$ is the rotation matrix (see Appendix
\ref{app:rotation} for an explicit definition). Following SW05 the
integral over $\hat{\bf R}$ can be performed analytically using the
relation
\be
\int d\hat{\bf R}\exp\left[-\rmi{\bk\cdot\br}({\bf R})\right]
=4\pi {\rm j}_0\left[kRf(\theta_k,\phi_k)\right]\ ,\label{Rhataverage}
\ee
where ${\rm j}_0$ is the zeroth order spherical Bessel function and
\be
f^2(\theta_k,\phi_k)=\cos^2\theta_k+\sin^2\theta_k
\left(\frac{a^2}{c^2}\cos^2\phi_k+\frac{b^2}{c^2}\sin^2\phi_k\!\!
\right).
\ee
Finally, on substituting equation (\ref{Rhataverage}) into
(\ref{window}) we find
\[ W_{i}(M)=\frac{(4\pi)}{8\pi^2}\int d\a \, p(\a|M)
 \frac{ab}{c^2} \int d\alpha\, d(\cos\beta)\, d\gamma \ ,\]
\be \hspace{1cm}\times\ \int
dR R^2 \, U(R,M) \, {\rm j}_0[k_i R f(\theta_{k_i}',\phi_{k_i}')]
\label{eq:w1}\ee
where we have from the rotation matrix that
\be \theta_k' = \arccos\left[ \frac{S_{\beta}\left( k_x
C_{\alpha}+k_y S_{\alpha}\right)+ C_{\beta} k_z}{k}
\right]\ ,\label{thetakprime}\ee
\be \phi_k' = \arccos\left[\frac{\eta}{k \sin{\theta_k'}}\right]\
,\ee
\be
\eta\equiv(C_{\beta}C_{\alpha}C_{\gamma}-S_{\alpha}S_{\gamma})k_x
+(C_{\beta}S_{\alpha}C_{\gamma}+C_{\alpha}S_{\gamma})k_y-S_{\beta}C_{\gamma}
k_z \label{phikprime} \ee
and where we have adopted the short-hand notation $C_{x}\equiv \cos x$
and $S_{x}\equiv\sin x$.  Importantly, if $a/c = b/c = 1$, then the
window function simplifies to that of the standard spherical halo
model.


\begin{figure}
\centerline{\includegraphics[width=8cm]{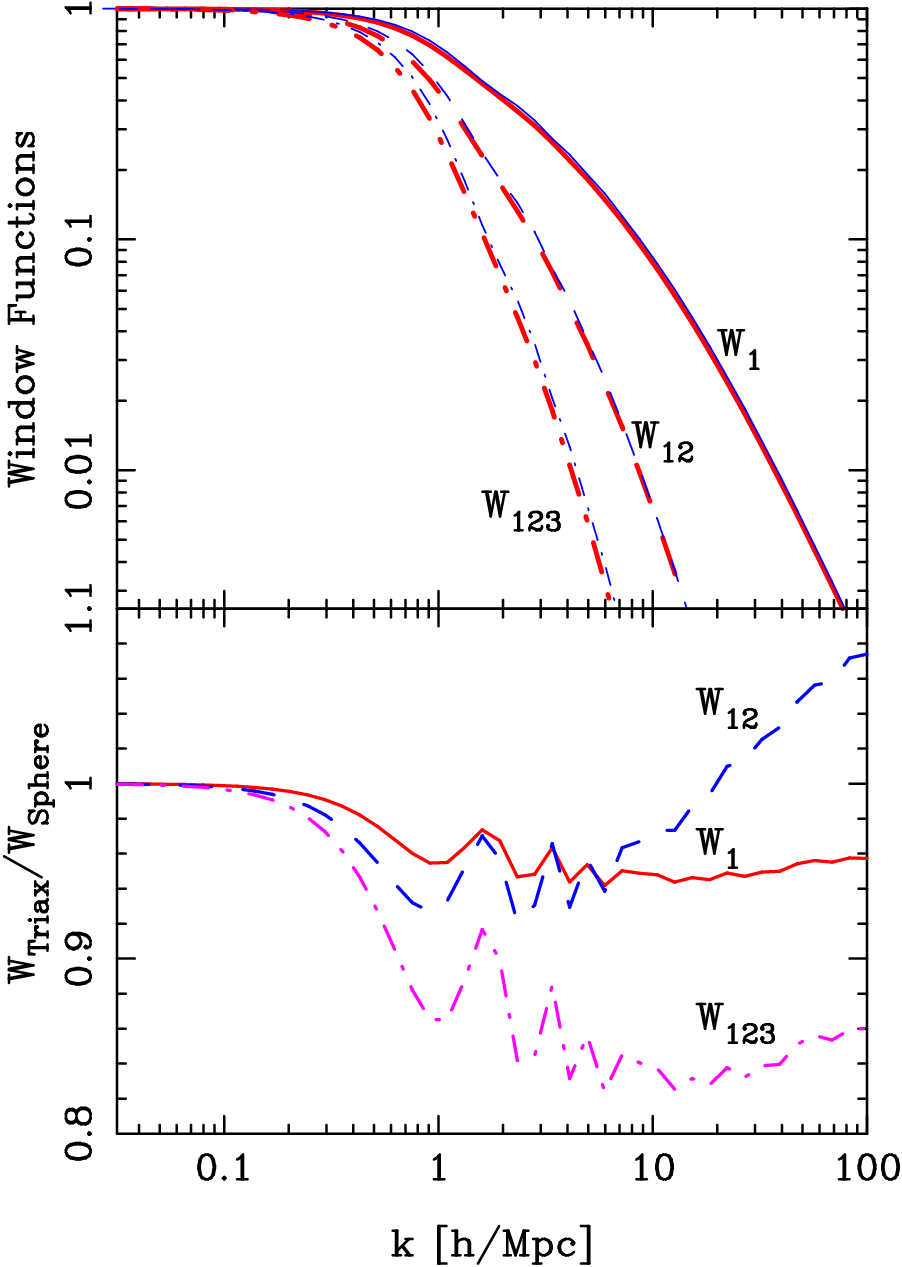}}
\caption{\small{The triaxial halo window functions compared with the
equivalent spherical halo window functions. In the top panel, thick
lines represent triaxial haloes and thin lines spherical.  Different
line styles represent $W_1$, $W_{12}$ and $W_{123}$ as defined by
equations (\ref{eq:w1}), (\ref{eq:w12}) and (\ref{eq:w123}). The lower
panel presents the ratio of the triaxial window functions to the
spherical.  The line styles are as in top panel.}
\label{fig:TriaxWindow}}
\end{figure}


Similarly, applying this procedure to the functions $W_{ij}$ and
$W_{123}$, we find:
\[
W_{ij}(M) \equiv \frac{(4\pi)^2}{8\pi^2}\int d\a \,
p(\a|M)\left(\frac{ab}{c^2}\right)^2 \int d\alpha\, d(\cos\beta)\,
d\gamma \]
\be \hspace{0.3cm} \times \prod_{\nu=\{i,j\}} \int
dR_{\nu} R_{\nu}^2 \, U(R_{\nu},M,\a) \, {\rm j}_0[k_{\nu} R_{\nu}
f (\theta_{k_{\nu}}',\phi_{k_{\nu}}')];\label{eq:w12}\ee
and
\[ W_{123}(M) \equiv \frac{(4\pi)^3}{8\pi^2}\int d\a \,
p(\a|M)\left(\frac{ab}{c^2}\right)^3 \int d\alpha\, d(\cos\beta)\,
d\gamma \]
\be \hspace{0.3cm}\times \prod_{\nu=\{1,2,3\}} \int
dR_{\nu} R_{\nu}^2 \, U(R_{\nu},M,\a) {\rm j}_0[k_{\nu} R_{\nu} f
(\theta_{k_{\nu}}',\phi_{k_{\nu}}')].
\label{eq:w123}\ee
In accordance with the set of conditions on the bispectrum, presented
in Section \ref{ssec:3ptGen}, we see that the window functions,
specified by equations (\ref{eq:w1}), (\ref{eq:w12}) and
(\ref{eq:w123}), are invariant under arbitrary rotations of the $k$
vector triple, and that they are also real functions of $k$.

As a point of practice for computing these functions, it is necessary
to specify the initial coordinates of the $k$-vector triple. There is
no loss of generality here, owing to the orientation average of the
halo. We therefore choose $\bk_1$ to be aligned with the $z$-axis of
the reference coordinate system. The orientation of $\bk_2$ is then
specified in spherical polar coordinates so that the polar angle gives
the angle between $\bk_1$ and $\bk_2$.  The azimuthal angle of $\bk_2$
is then, from equations (\ref{thetakprime}) and (\ref{phikprime}),
entirely arbitrary. Once $\bk_2$ has been specified, $\bk_3$ is then
fixed through the triangle constraint. The window functions, and
subsequently the bispectrum, then simply become functions of
$|\bk_1|$, $|\bk_2|$ and the angle $\theta = \arccos\left(
\bk_1\cdot\bk_2/k_1k_2\right)$, as required for homogeneous and
isotropic random fields.

Figure \ref{fig:TriaxWindow} shows a calculation of the window
functions given by equations (\ref{eq:w1}), (\ref{eq:w12}) and
(\ref{eq:w123}) for equilateral triangle configurations, i.e $|\bk_1|
= |\bk_2| = k$ and $\theta_{12} = 2\pi/3$. In this calculation we have
taken the ellipsoidal density profile to be the triaxial NFW model
(see Section \ref{ssec:details}), and for illustration we have taken
the joint axis ratio PDF to be simply the product of two delta
functions: $p(a/c,b/c)=\delta^D\!(a/c-0.5) \, \delta^D\!(b/c-0.5)$.
This constrains all haloes to take the form of prolate ellipsoids. The
top panel shows the two usual characteristics: on large scales the
window function approaches unity and on small scales it rapidly decays
with oscillatory features present. To show the effect of halo
triaxiality on these window functions in more detail, we compute the
ratio of these quantities with the windows for the spherical model
that result when $p(a/c,b/c)=\delta^D\!(a/c-1) \,
\delta^D\!(b/c-1)$. The results of this are presented in the lower
panel of the same figure. Clearly, the function $W_{123}$ is much more
sensitive to the shape of the haloes than either $W_1$ or
$W_{12}$. This leads us to conclude that the 1-Halo term becomes
increasingly sensitive to the halo shape as the order of the
clustering statistic increases.


\begin{figure*}
\centerline{
  \includegraphics[width=8.5cm]{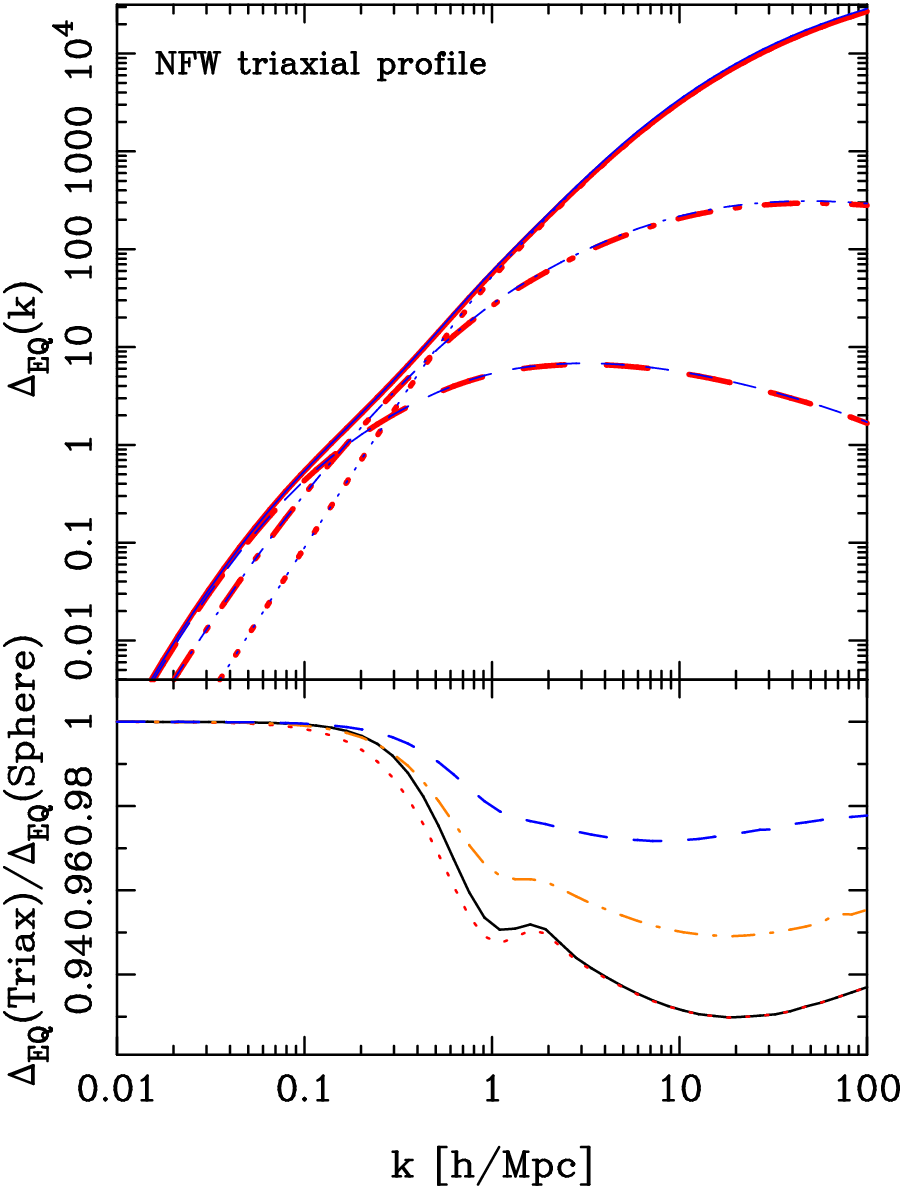}
  \includegraphics[width=8.5cm]{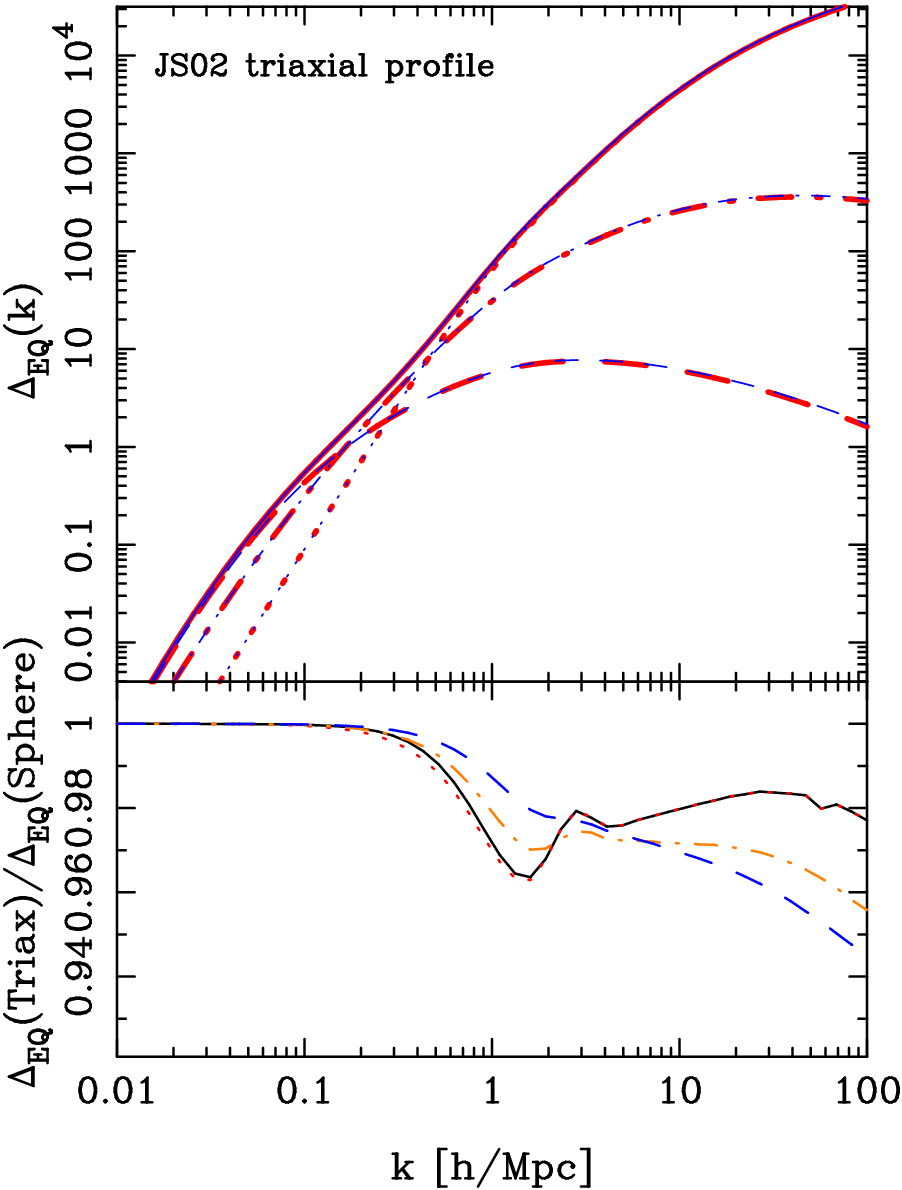}
}
\caption{\small{The equilateral bispectrum. Top panel presents the
dimensionless bispectrum as defined in equation (\ref{eq:BiEq}). Thick
lines represent the triaxial haloes and thin lines the equivalent
spherical haloes. The dot, dot-dash and dash lines represent the 1-,
2- and 3-Halo terms respectively and the solid lines represent the
total. Bottom panel presents the ratio of the triaxial halo bispectrum
to the spherical. The line styles are preserved, but note that the
dotted line represents the ratio of the triaxial 1-Halo term to the
spherical halo model 1-Halo term, similarly for the 2- and 3-Halo
terms.}
\label{fig:bispecEQ}}
\end{figure*}


\section{Results: Mass bispectrum}\label{sec:results}

In this section we evaluate the bispectrum given by the summation of
equations (\ref{eq:bi1H}), (\ref{eq:bi2H}) and (\ref{eq:bi3H}), using
our simplified expressions for the window functions given by equations
(\ref{eq:w1}), (\ref{eq:w12}) and (\ref{eq:w123}).


\subsection{Computational details}\label{ssec:details}

We investigate two models for the halo profile $U(R,M,\a)$: the first
is a toy model that allows us to explore how modification of the halo
shape alone affects the clustering -- we refer to this as the
`triaxial NFW' model; the second is the more realistic density profile
model of JS02, in which halo shape and central densities are not
independent -- we refer to this as the `JS02 model'. For a more
thorough explanation of these we refer the reader to SW05 and JS02,
respectively; note that in SW05 the triaxial NFW model is referred to
as the `continuity model'. Common to both of these models is the
density run, which is assumed to follow the NFW form
%
\be \rho(R) = \frac{\delta_c^{\rm triax}(M) \bar{\rho} }{y(1+y)^2};
\,
\hspace{10mm} y\equiv R/R_0(M),
\ee
where $R$ is the ellipsoidal radial parameter from equation
(\ref{eq:isodef}), $\delta_c^{\rm triax}(M)$ is the characteristic
density of the halo and $R_0(M)$ is the scale radius; in the JS02
model these are treated as independent variables.

The PDF for the axis ratios, $p(\a|M)$, is taken from JS02. We point
out that in a recent paper, \citet{LeeJingSuto2005} valiantly
attempted to derive the halo axis ratio PDF from the peak theory of
Gaussian random fields.  However, their predictions appear to be in
conflict with the results from $N$-body simulations and so we do not
consider their model further. We use the \citet{ShethTormen1999} mass
function and halo bias functions, and correct for the different
definitions of halo mass as discussed in SW05.  In order to focus
purely on the effects of triaxial modelling, we do not draw the
concentration parameter from a probability distribution, as advocated
by JS02, but treat it as a deterministic variable related to the halo
mass.

Although in the previous section we greatly simplified our expressions
for the halo window functions, we are still faced with the challenge
of performing 7-D numerical integrals to compute the bispectrum.
Evaluation of these expressions through a set of serial quadratures is
unfeasible on a standard workstation. Instead, we advocate the use of
an efficient multi-dimensional algorithm, such as the Korobov-Conroy
\citep{Korobov1963,Conroy1967} or Sag-Szekeres algorithm
\citep{SagSzekeres1964}. However, the integrals over the ellipsoidal
radial parameter $R$ require more care, due to the oscillatory nature
of the integrand. To evaluate these we therefore employ a 1-D adaptive
routine. Thus, for example, to evaluate the 1-Halo contribution to the
bispectrum for a given $k$-space triangle (equation \ref{eq:bi1H}), we
compute a 6-D numerical integral of a function that is itself the
product of three 1-D integrals. Such numerical computations are
achievable on a modern PC in a reasonable time frame. In Section
\ref{ssec:robust} we present an independent determination of the
triaxial halo model bispectrum, obtained through direct measurement
from synthetic halo simulations. This provides a robust test of the
veracity of the results produced by the multi-dimensional integrators.


\begin{figure*}
\centerline{ 
\includegraphics[width=7.0cm]{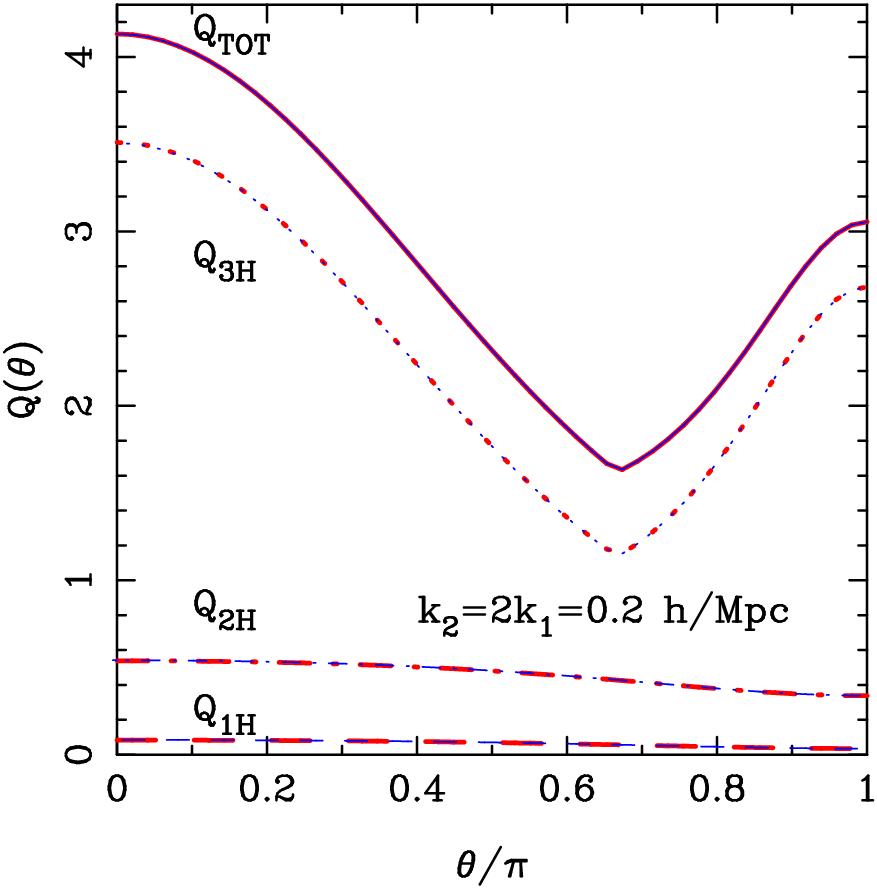}
\includegraphics[width=7.0cm]{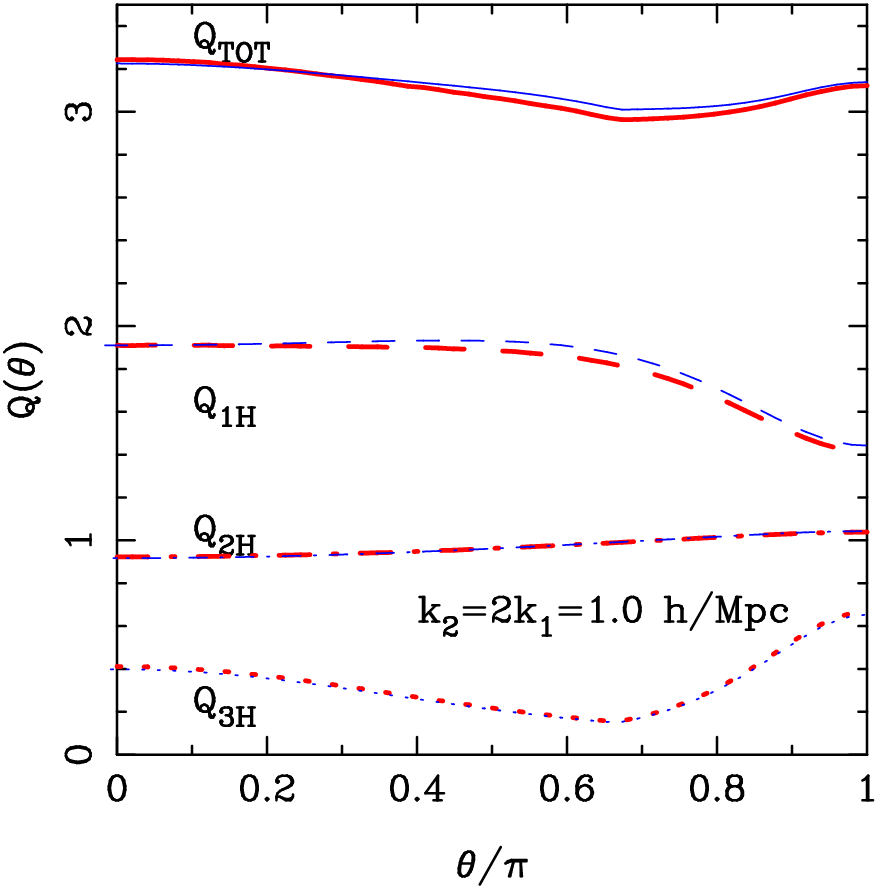}}
\centerline{
\includegraphics[width=7.0cm]{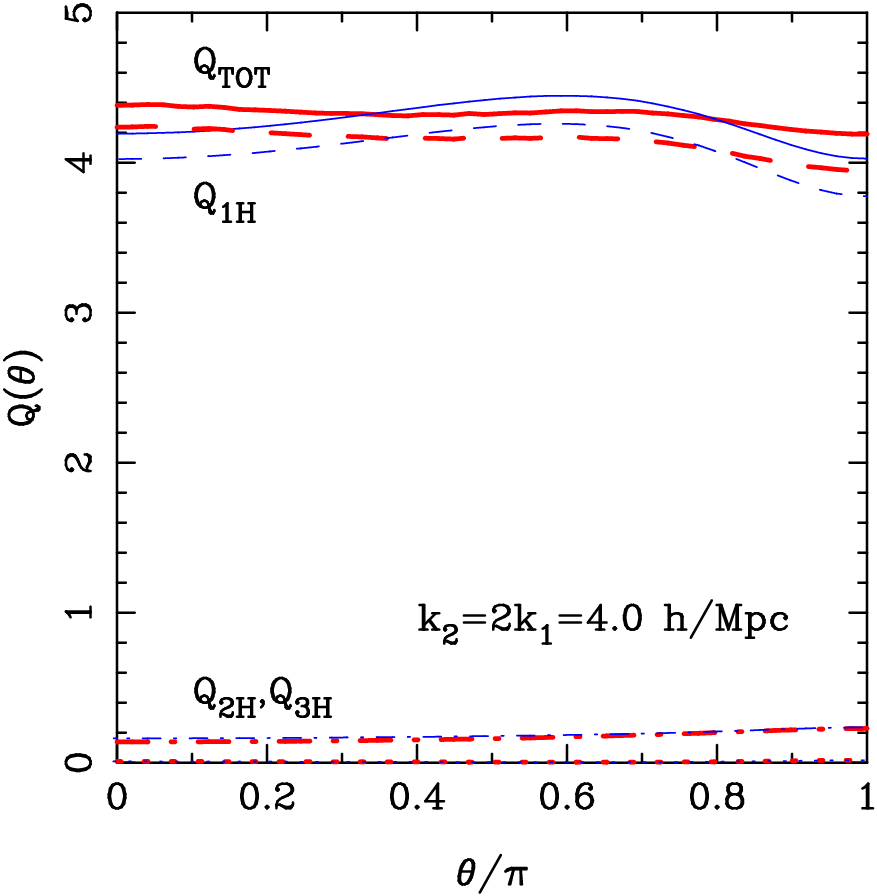}
\includegraphics[width=7.0cm]{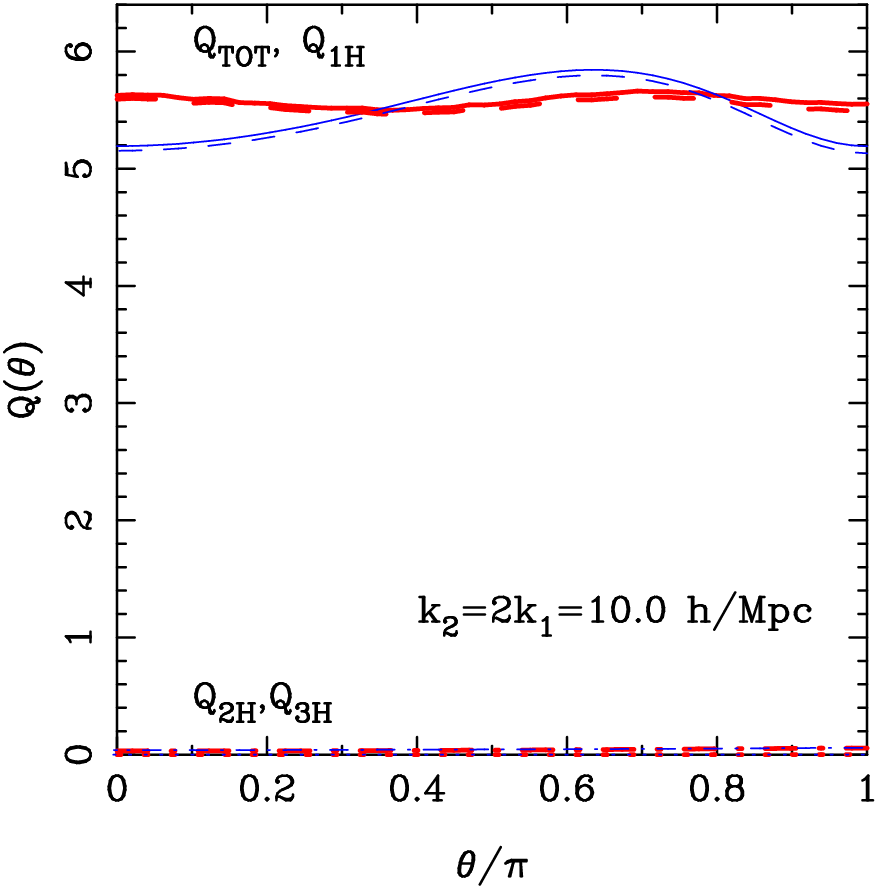}}
\caption{\small{Reduced bispectrum $Q$ as a function of
$\cos\theta\equiv\bk_1\cdot\bk_2/k_1k_2$, where $Q$ is defined through
equation (\ref{eq:Q}) and where we have considered the particular
triangle configuration $|k_2|=2|k_1|$. The six panels show how the
$Q$--$\theta$ curves evolve from large linear scales to small
nonlinear scales; the sequence is shown for
$k_2=(0.2,1.0,4.0,10.0)\,h\,{\rm Mpc}^{-1}$. Thick lines show the
predictions of the triaxial halo model, thin lines are those for
spherical haloes.}
\label{fig:Qtheta}}
\end{figure*}


\subsection{Equilateral triangles}

Figure \ref{fig:bispecEQ} (top panels) shows the predictions for the
dimensionless bispectrum for equilateral triangles and for the
triaxial NFW and JS02 models. This is defined as
\be \Delta_{\rm EQ}(k)=\frac{4\pi}{(2\pi)^3} k^3 \sqrt{B_{\rm
EQ}(k)}\label{eq:BiEq}\ ,\ee
where $B_{\rm EQ}(k)$ is the bispectrum for equilateral triangles. In
addition to results for triaxial haloes (thick solid lines), we also
show the results for spherical haloes (thin lines). The spherical halo
model predictions are obtained by setting the axis ratio PDF to be a
product of delta functions as discussed in Section
\ref{ssec:orientation}. It is evident that deviations from the
spherical halo model appear to be very small for both profiles. In the
bottom panels of Figure \ref{fig:bispecEQ}, we illustrate the
differences more clearly by plotting the ratio $\Delta^2_{\rm eq}({\rm
triaxial})/ \Delta^2_{\rm eq}({\rm spherical)}$ for the 1-, 2- and
3-Halo term, and their sum, respectively. As may be expected from the
window functions, it is clear that the biggest departure from the
spherical model occurs in the 1-Halo term, which is at most $\sim$
7\%. The 2- and 3-Halo terms also deviate by several percent, though
only on scales below which their contribution to the total bispectrum
is insignificant. On larger scales there is no appreciable difference
from the spherical model.  Thus, for the purposes of halo model
calculations that wish to measure the equilateral triangle bispectrum,
it is a good approximation to neglect the halo shape information in
the 2- and 3-Halo terms.  However, if one wishes to be accurate to the
level of $\sim5\%$ one must include shape information on the 1-Halo
term.

On the other hand, it is now clear that the discrepancies ($\sim
50\%$) between the measurements of the \emph{equilateral triangle}
bispectrum from numerical simulations and the predictions for the halo
model, noted in the work of SSHJ and \citet{Fosalbaetal2005}, can not
solely be attributed to the break down of the spherical halo
approximation.


\begin{figure*}
\centerline{
\includegraphics[width=8.0cm]{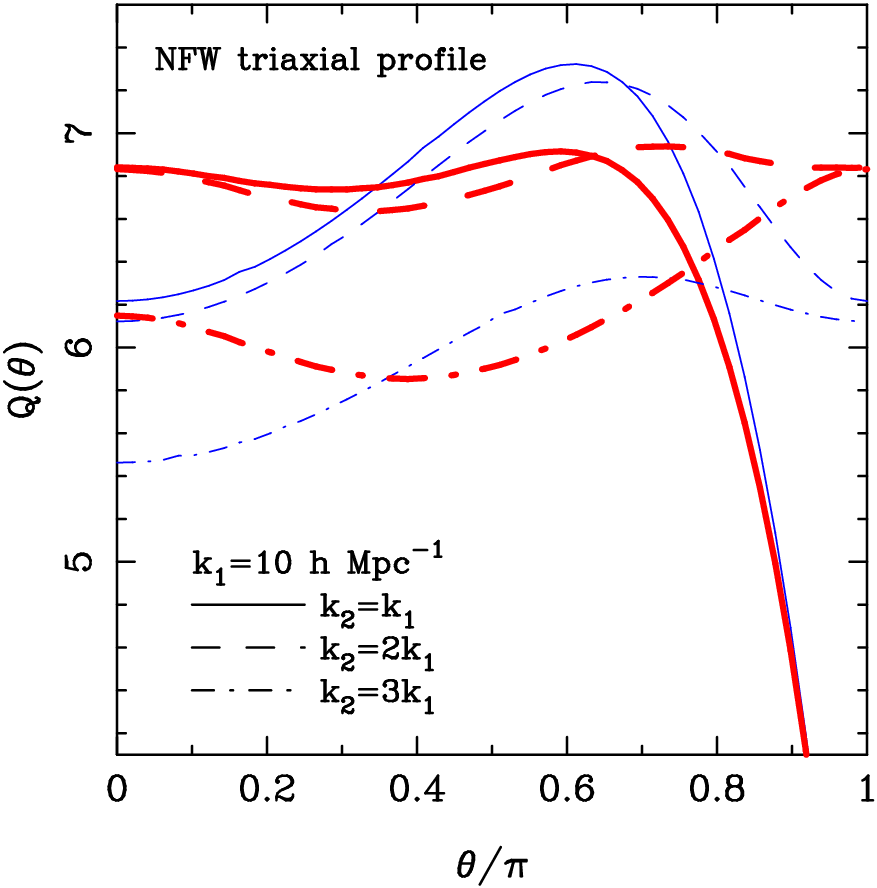}
\includegraphics[width=8.0cm]{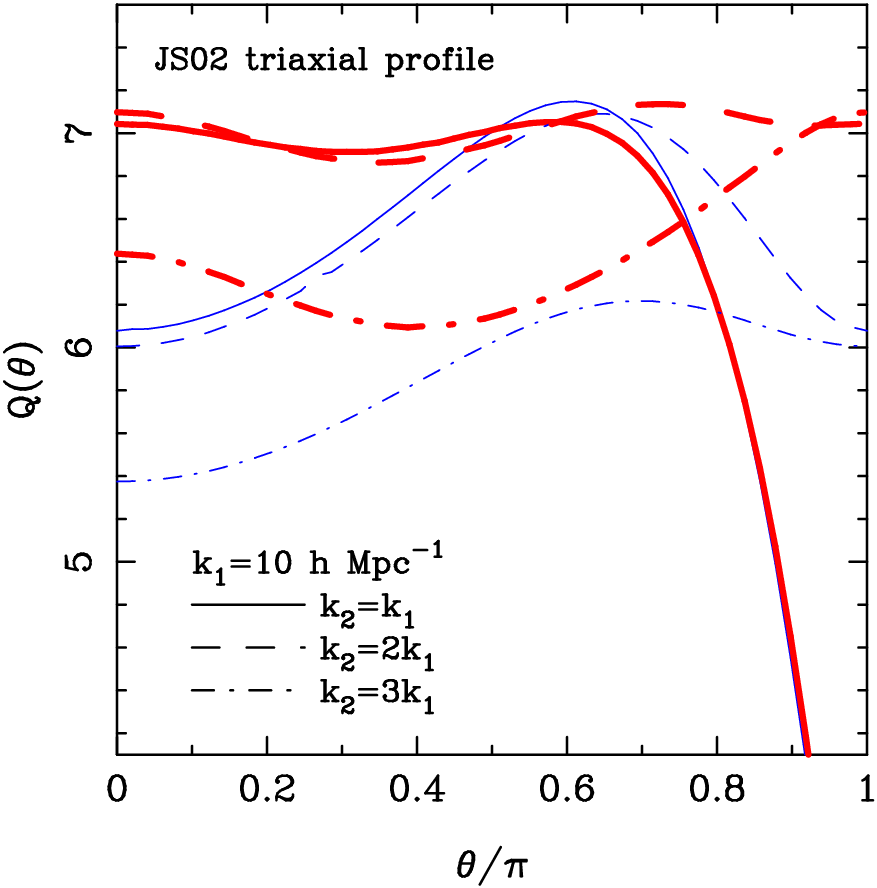}}
\caption{\small{Dependence of $Q(\theta)$ on the ratio
    $k_2/k_1$. Thick lines represent the triaxial halo model
    predictions and thin lines denote the spherical model.  Solid,
    dash and dot-dash lines represent the cases where the triangle
    configurations: $k_2=k_1$; $k_2=2k_1$; $k_2=3k_1$.}
\label{fig:Qtheta2}}
\end{figure*}


\subsection{Variation with triangle configuration}

We next consider the dependence of the bispectrum on the configuration
of Fourier space triangles. This is achieved by fixing the lengths of
$k_1$ and $k_2$ and then varying the angle between them.  Since the
effects caused by halo triaxiality are relatively small, we choose to
examine the more sensitive reduced bispectrum, or hierarchical
amplitude:
\be Q_{123}\equiv
\frac{B_{123}}{P(k_1)P(k_2)+P(k_2)P(k_3)+P(k_3)P(k_1)}
\label{eq:Q},
\ee
where, for the case of the triaxial model predictions, $P(k)$ is the
triaxial halo model power spectrum (SW05). In Figure \ref{fig:Qtheta}
we show the results for the triaxial NFW profile model on various
scales.  In all four panels we have taken one of the fixed sides of
the triangle to be twice the length of the other fixed side:
$k_2=2k_1$. The four panels show the results for the case where
$k_2=(0.2,1.0,4.0,10.0)\,h\,{\rm Mpc}^{-1}$.

The generic features of these curves have been examined in great
detail in previous studies (Scoccimarro et al. 1998; SSHJ; Hou et
al. 2005; Fosalba, Pan \& Szapudi 2005).  On large scales, it has been
suggested that the variation in the configuration dependence can be
attributed to coherent filamentary structures in the density field
enhancing the contribution to $Q$ from collinear triangles
\citep{Scoccimarroetal1998}. On intermediate scales, the density field
begins to be dominated by collapsing objects that have almost
isotropic spatial distributions, thus $Q$ shows much reduced variation
across the configuration. Finally, on small scales, $Q$ is completely
determined by the structure and shapes of the dark matter haloes. This
overall evolution is well reproduced in our Figure \ref{fig:Qtheta}.

Considering the relative differences between the triaxial and
spherical predictions in Fig. \ref{fig:Qtheta}, we make the following
observations: on large scales ($k_2 = 0.2 \,h\,{\rm Mpc}^{-1}$; top
left panel), where $Q$ is dominated by the 3-Halo term, the affect of
halo triaxiality is insignificant. This can easily be understood
through considering the $k\rightarrow0$ limit of equation
(\ref{eq:w1}); On intermediate scales ($k_2 = 1 \,h\,{\rm Mpc}^{-1}$;
top right panel), the magnitudes of the 1-, 2- and 3-Halo terms are
comparable and one sees that there is a very small relative difference
between the spherical and triaxial halo model predictions.  This
difference originates from the growing 1-Halo term, there being no
perceivable variations in the 2- or 3-Halo terms; On smaller scales
(lower panels), $Q$ is completely dominated by the 1-Halo term, and we
see that there is a significant difference between the spherical and
triaxial halo model predictions.  The variation in $Q$ across the set
of triangle configurations, for the triaxial haloes, appears to be
much flatter than the variation apparent in the spherical halo
model. It should also be noted that the positions of the maxima and
minima of $Q$ are shifted.


\begin{figure*}
\centerline{
\includegraphics[width=7.0cm]{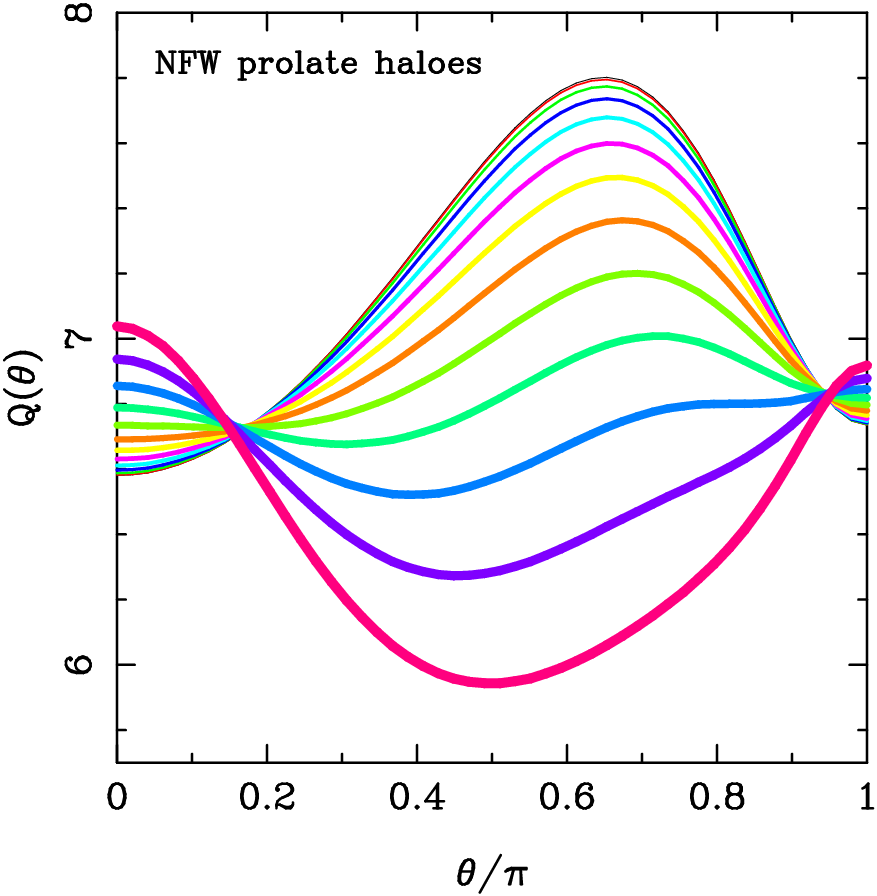}
\includegraphics[width=7.0cm]{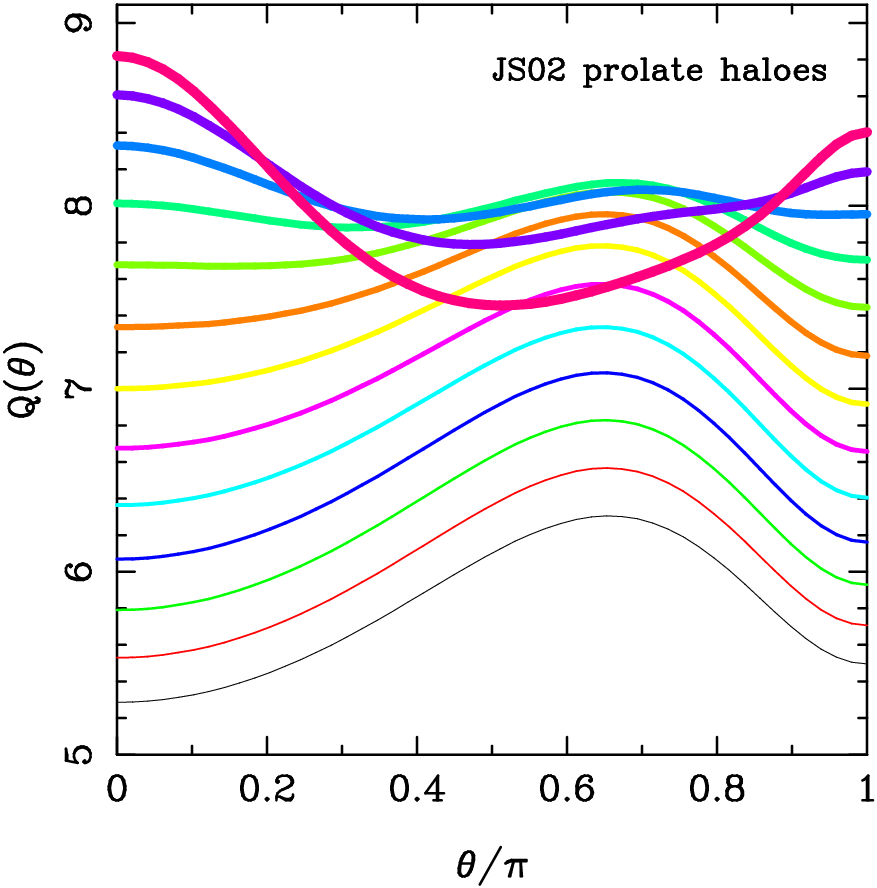}}
\caption{\small{Dependence of $Q(\theta)$ on halo prolaticity.  The
axis ratio PDF is taken to be a product of delta functions. Increasing
line thickness corresponds to increasing halo prolaticity, with the
thinest line being $a/c=b/c=1.0$ and the thickest being
$a/c=b/c=0.4$.}
\label{fig:Qtheta3}}
\end{figure*}


\begin{figure*}
\centerline{
\includegraphics[width=7.0cm]{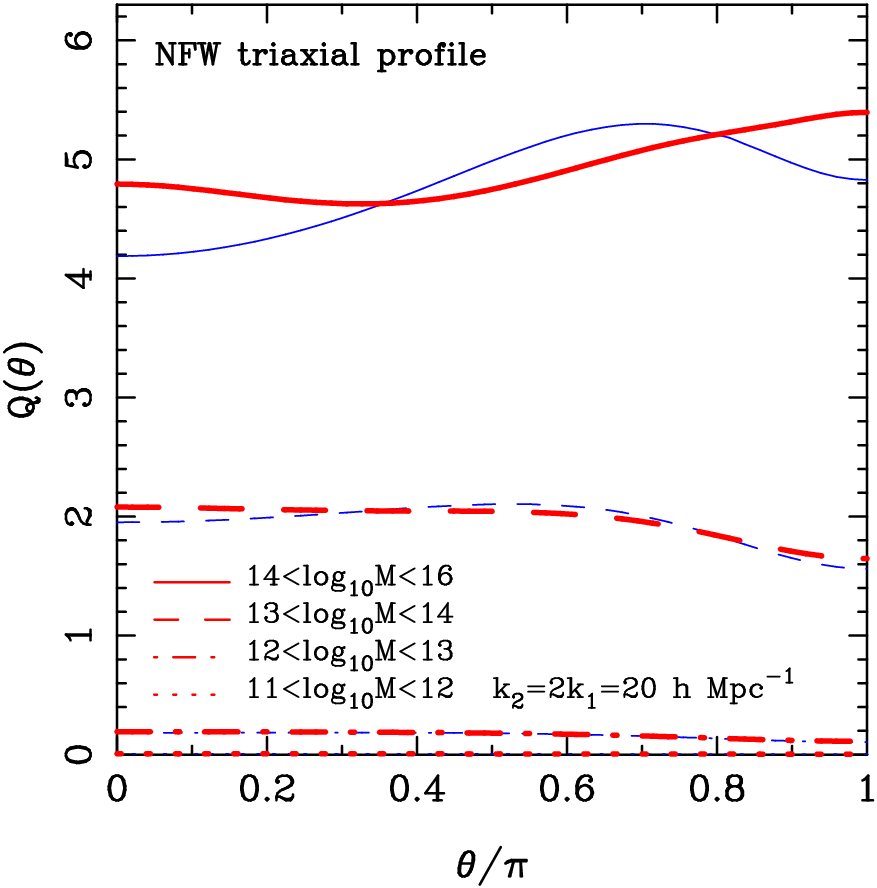}
\includegraphics[width=7.0cm]{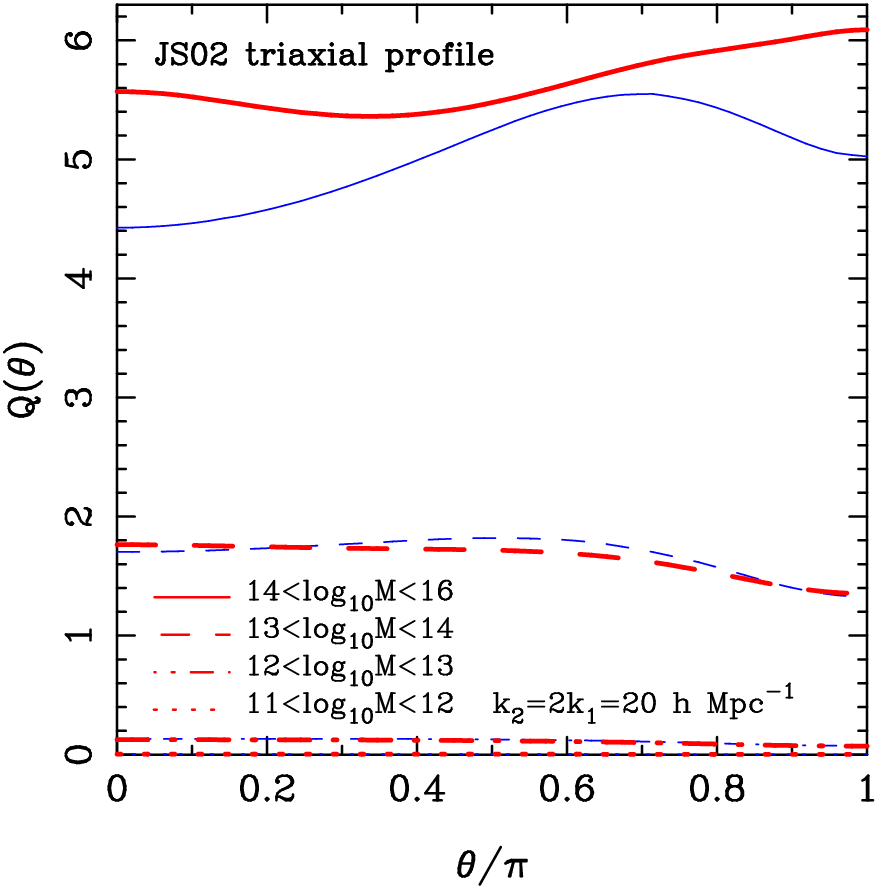}}
\caption{\small{Dependence of $Q(\theta)$ on halo mass. Again, thick
lines represent triaxial model predictions thin lines spherical.  }
\label{fig:Qtheta4}}
\end{figure*}


In Figure \ref{fig:Qtheta2} we investigate these apparent differences
between the spherical and triaxial halo model in closer detail. Fixing
$k_1 = 10 \, h\,{\rm Mpc}^{-1}$, we consider the cases where
$k_2/k_1=(1,2,3)$. Further, since the 1-Halo term dominates these
scales, we focus in on the region over which it varies and ignore all
other terms. In all cases we see that the triaxial model predictions
for $Q$, for both density profile models, are suppressed with respect
to the spherical model on angular scales where the $k$-vectors are
close to equilateral triangles, and amplified on scales where the
$k$-vectors are colinear and nonvanishing. Most importantly, from the
figures it is clear that the overall functional forms exhibited by the
predictions depend little on the density profile model, modulo an
amplitude shift, and that it is the shape of the haloes that dictates
the shape of the curves.

In the work of SSHJ, it was demonstrated that on small scales the
predictions for the spherical halo model for $Q(\theta)$ exhibited a
convex configuration dependence, whereas the measurements from N-body
simulations showed a much flatter dependence with $\theta$.  Indeed,
SSHJ speculated that this disagreement was likely the result of a
breakdown in the assumption of halo sphericity. More recently these
discrepancies have been highlighted by the work of
\citet{Fosalbaetal2005}, who demonstrated that even with a somewhat
\emph{ad-hoc} modification of the halo boundaries, the configuration
dependence on small scales can not be replicated in the spherical halo
model. On contrasting these results with those above, we are lead to
believe that the apparent discrepancies between theory and numerical
simulation are likely attributed, in the main, to a break down in the
spherical halo approximation. Moreover, it is highly unlikely that the
exhibited configuration behaviour could be reproduced by other
variations in the halo modelling prescription, e.g. through the
stochasticity of the halo concentration parameter or through the
inclusion of substructures. Such changes may alter the amplitude of
$Q(\theta)$, as between the left and right panels of Figure
\ref{fig:Qtheta2}, but they should not change the overall shape.


\subsection{Dependence on halo shape}

We now explore in closer detail how sensitive the triaxial halo model
predictions are to the precise choice of the PDF for the axis ratios.
This we do most directly by taking $p(\a|M)$ as a product of Dirac
delta functions (see Section \ref{ssec:orientation}).  The sequence of
curves in Figure \ref{fig:Qtheta3} shows, in order of increasing
thickness, the predictions for prolate dark matter haloes with axis
ratios in the range $0.4 \le a/c = b/c \le 1.0$, incremented in steps of
$a/c=b/c=0.05$.  Again we have set $k_2/k_1 = 2$ and $k_1 = 10
\,h\,{\rm Mpc}^{-1}$. For both the triaxial NFW profile and the JS02
model a generic trend is apparent: for spherical haloes, the maximum
in $Q(\theta)$ occurs for $\theta = 2\pi/3$, with minima at $\theta =
0$ and $\pi$. As haloes become more ellipsoidal the maxima in
$Q(\theta)$ is strongly suppressed while the minima are enhanced. For
the most prolate objects the maxima in $Q(\theta)$ lie at $\theta = 0$
and $\pi$, while a deep minimum forms between $\theta = \pi/3$ and
$2\pi/3$.

These trends can readily be understood by considering all possible
placements of the $k$-space triangle within the Fourier transformed
halo profile $U(\bk_i,M,\a,\avec)$. Varying the configuration of the
triangle, one sees that the largest contribution to the bispectrum
will come from triangles that most optimally fill the halo volume in
$k$-space. For a triaxial ellipsoid, its Fourier transform is also
ellipsoidally symmetric. Thus, the `optimal' triangle configuration is
the one that matches most closely the halo's symmetry. For spherical
haloes this clearly means equilateral triangles (or close to); whereas
for very ellipsoidal objects (e.g. the thickest lines in
Fig. \ref{fig:Qtheta3}), the optimum configurations shift to those
triangles where the $k$-vectors are coaligned. This explanation
applies equally well to both the NFW and JS02 profiles since, by
construction, they both share the same axis ratio PDFs. The apparent
amplitude shifts in the JS02 predictions for $Q(\theta)$ are due to
the fact that in this model, both the central density and
concentration parameter depend on the halo axis ratios.


\subsection{Dependence on halo mass}

In the numerical simulation work of JS02 it was found that, for CDM
models, the axis ratio PDF was conditional upon halo mass, and that
more massive haloes were on average more triaxial than lower mass
haloes. This can be understood from halo formation histories: high
mass haloes form at late times in the Universe, whereas low mass
haloes form at early times.  Therefore, lower mass haloes have a
longer period of time for particle orbits to undergo relaxation and
become isotropic. Furthermore, since the haloes are assembled through
the anisotropic accretion of lower mass haloes flowing along
filaments, the halo shape at the time of formation will be more
closely described as a triaxial ellipsoid. It is therefore of interest
to consider how the bispectrum predictions depend on the abundances
and masses of haloes.

Figure \ref{fig:Qtheta4} shows the contribution to $Q(\theta)$ arising
from four coarse bins in halo mass, for both the triaxial NFW and JS02
density profile models. Once again we set $k_2/k_1 = 2$ and $k_2 = 20
h\,{\rm Mpc}^{-1}$. We take $P(k)$ in equation (\ref{eq:Q}) as the
power due to {\em all} halo masses, so that the sum of the
contributions from the four bins equals the total $Q$. The main
contribution to $Q$ comes from haloes with masses in the range
$10^{14} h^{-1} M_{\sun} < M < 10^{16} h^{-1}M_{\sun}$. We see that
both the triaxial and spherical model predictions exhibit a
`wave-like' form with $\theta$. These waves are of similar amplitudes,
but exhibit different `phases'. The maximum amplitude for the triaxial
halo model predictions occurs at $\theta = \pi$, with a second peak of
slightly lower amplitude at $\theta=0$, whereas for the spherical
model, the peak amplitude is located at $\theta=2\pi/3$.  These
differences can be understood from the discussion surrounding Figure
\ref{fig:Qtheta3}.

The next most significant contribution comes from haloes with masses
in the range $10^{13} M_{\sun}h^{-1} < M < 10^{14}
M_{\sun}h^{-1}$. These curves show less of the wave-like form
exhibited in the higher mass haloes. Crucially, however, for the case
of the spherical haloes the maxima and minima in $Q$ occur at the same
position as for the higher mass bin. For the triaxial predictions this
is not the case, and the form of the curve has become almost
consistent with that of the spherical halo model. This can be
understood through the shifting of the mean of the JS02 halo axis
ratio PDF to more spherical haloes as the halo mass
decreases. Finally, one can now see how for the triaxial halo
predictions the summation of the contributions to $Q$ from high and
low mass haloes add de-constructively to give a relatively flat
function across $\theta$ (cf. Fig. \ref{fig:Qtheta}).


\begin{figure}
\centerline{
\includegraphics[width=8cm]{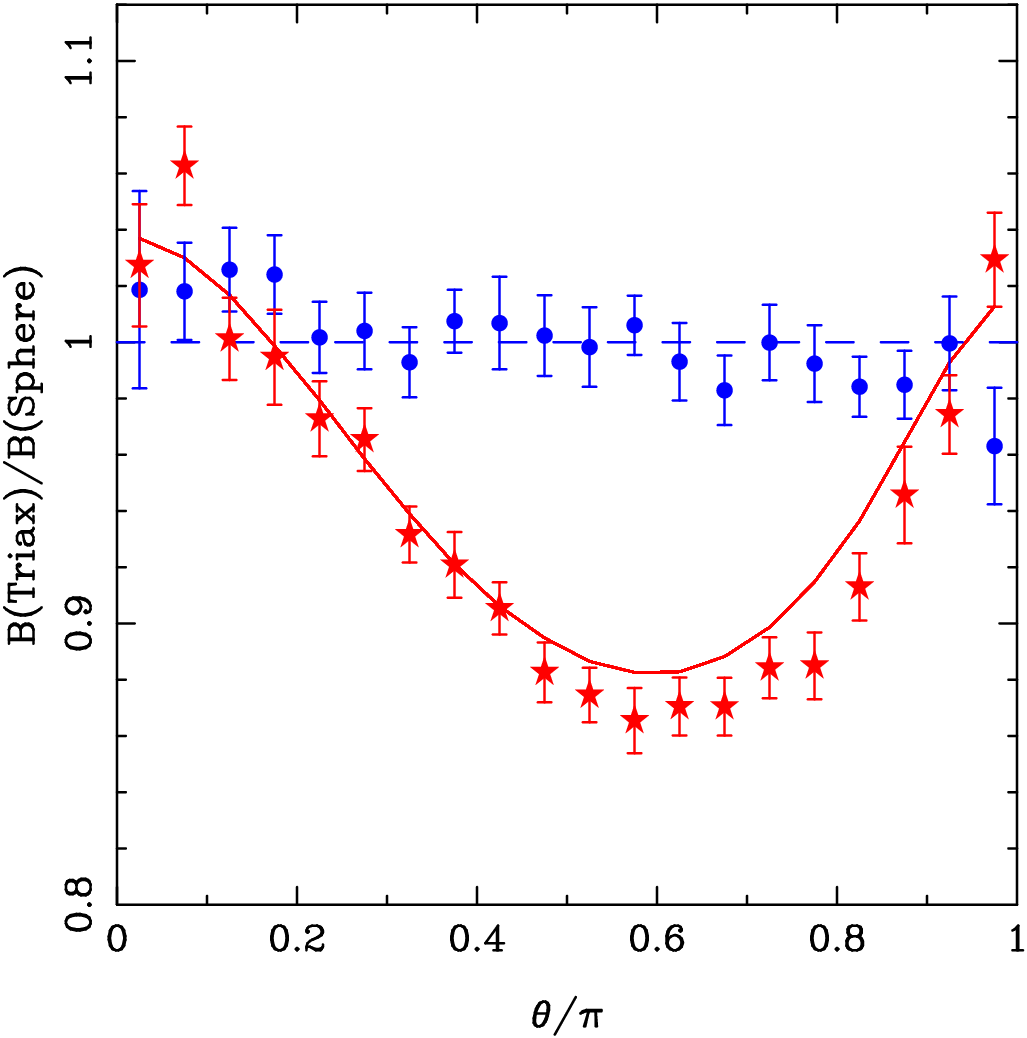}}
\caption{\small{Test of the robustness of predictions of the triangle
configuration dependent bispectrum. The star points show the mean
bispectrum measured from an ensemble of 20 synthetic triaxial halo
density fields ratioed to the predictions from the analytic spherical
halo model. The circular points show the same, but for synthetic
spherical halo density fields. The solid line shows the analytic
predictions for the triaxial halo model.}
\label{fig:QRobust}}
\end{figure}


\subsection{Robustness of predictions}\label{ssec:robust}

Owing to the relative obscurity of the high-dimensional integrators
that we have used to solve the integrals in Section
\ref{ssec:bispectrum}, we feel that it is necessary to demonstrate the
robustness of their predictions. This we accomplish through the use of
synthetic halo simulations. Following \citet{PeacockSmith2000} and
\citet{ScoccimarroSheth2002}, we generate direct realizations of the
triaxial halo density field through populating cubical regions of
space with haloes that are sampled from the mass function, down to
some limiting mass. Within each halo we then distribute equal mass
particles in accordance with the required density structure, until the
halo mass is reached. The spatial positions of the halo centres are
chosen at random so that there is no large scale clustering of
halos. From this particle distribution we then directly measure the
bispectrum.

Figure \ref{fig:QRobust} shows the measurement of the triaxial halo
model and spherical halo model bispectrum in an ensemble of twenty
synthetic simulations, ratioed with the analytic predictions of the
spherical halo model. These are then compared with the numerically
integrated predictions of our analytic results for the bispectrum
1-Halo term. We looked at triangles for which $k_2 = 2 k_1 = 20
h^{-1}$ Mpc. Clearly, the analytic predictions and direct measurements
from the synthetic data are in excellent agreement (2$\sigma$
level). We are therefore confident that, for these problems, the
integrators are accurate to at least $\sim1\%$.

To summarize the main results of Section \ref{sec:results}, we have
convincingly demonstrated that the detailed shape of $Q(\theta)$ on
small scales depends very sensitively on the shapes of dark matter
haloes, and also their axis ratio PDF. Furthermore, we propose that
the discrepancies observed between the theoretical predictions for
$Q(\theta)$ and direct measurements from $N$-body simulations may be
largely explained through the break down in the spherical halo
approximation.


\section{Galaxy clustering}\label{sec:galaxies}

We now turn our attention to the second of the questions posed in the
introduction, and explore to what extent halo shapes impact on the
galaxy clustering.  Such information is of great importance if one
wishes to use the halo model as a means for constraining the
parameters of the HOD from observations.


\subsection{Galaxies in the halo model}\label{ssec:galhalomodel}

As a number of authors have shown, the halo model formalism can be
extended to successfully model the clustering of galaxies (Seljak
2000; Peacock \& Smith 2000; SSHJ; Berlind \& Weinberg 2002, Yang et
al. 2003, Cooray \& Sheth 2003).  The methodology follows ideas first
laid down in earlier works by \cite{NeymanScott1952} and others, but
which are nicely summarized in \cite{Peebles1980}.

We note at the outset that one can write down the hierarchy of
correlation functions directly from the mass clustering statistics by
making use of the following simple substitutions.  Firstly, we make
the transformation
\be U(\br,M,\a,\avec)\rightarrow U_g(\br,M,\a,\avec)\equiv\frac{\rho_g(\br)}
{\int d\br \rho_g(\br)}\ ,\ee
where $U_g$ describes the average spatial distribution of galaxies
within each halo normalized by the expected number of galaxies.  Note
that a sub- or superscript $g$ labels galaxy quantities. Secondly,
instead of weighting the integrals over the halo mass function by
mass, we weight by either the expected number of galaxies, or galaxy
pairs, triplets etc., depending on whether we are considering the 3-,
2- or 1-Halo terms respectively. Thus we have
\ba
n(M)M & \rightarrow & n(M)\left<N|M\right>\ ,\\
n(M)M^2 & \rightarrow & n(M)\left<N(N-1)|M\right>\ ,\\
n(M)M^3 & \rightarrow & n(M)\left<N(N-1)(N-2)|M\right>\ ,
\ea
where
\be \left<N|M\right>=\sum_{N=1}^{\infty}N\, P(N|M)\ ,\ee \be
\left<N(N-1)|M\right>=\sum_{N=2}^{\infty}N(N-1)\, P(N|M)\ ,\ee \be
\left<N(N-1)(N-2)|M\right>=\sum_{N=3}^{\infty}N(N-1)(N-2)\, P(N|M)\ .
\ee
In the above equations $P(N|M)$ is the conditional halo occupation
probability that gives the probability of finding $N$ galaxies in a
halo of mass $M$.  Lastly, in order to correctly normalize the
clustering we perform the transformation
\be \rhob\rightarrow \nbarg = \int dM\,n(M)\left<N|M\right>\ ,\ee
where $\nbarg$ is the mean number density of galaxies.


\subsection{Galaxy correlation functions}

Following the procedure described in the previous section, we can
write down the two- and three-point correlation functions for the
galaxies. Thus, for the two-point function, we find $\xi_{\rm g}$ to
be:
\be \xi_{\rm g}=\xi^{\rm 1H}_{\rm g}+\xi^{\rm 2H}_{\rm g}\ ;\ee
\[ \xi^{\rm 1H}_{\rm g}=\frac{1}{\nbarg^2\,8\pi^2} \int dM \,
d \bx\, d\a \, d\avec \,n(M) \left<N(N-1)|M\right> p(\a|M) \] \be
\hspace{1cm}\times\ \prod_{i=1}^{2} \left\{U_{\rm
g}(M,\br_i-\bx,\a,\avec)\right\}\ ;\ee
\[ \xi^{\rm 2H}_{\rm g} = \frac{1}{(\nbarg\,8\pi^2)^2} \int
\prod_{i=1}^{2} \left\{ dM_i \,d \bx_i\, d\a_i\, d\avec_i\,
n(M_i)\, \left<N|M_i\right> \right. \] \be \hspace{1cm}\times \,
\left. p(\a_i|M_i) \, U_{\rm
g}(M_1,\br_1-\bx_1,\a_1,\avec_1)\right\} \xi^s_2(1,2)\ .\ee
Similarly, we find $\zeta_{\rm g}$ to be:
\be \zeta_{\rm g}=\zeta^{\rm 1H}_{\rm g}+\zeta^{\rm 2H}_{\rm
g}+\zeta^{\rm 3H}_{\rm g}\ ;\ee
\[ \zeta^{\rm 1H}_{\rm g}=\frac{1}{\nbarg^3 8\pi^2}
\int dM \,d \bx \,d\a \, d\avec \, n(M)
\left<N(N-1)(N-2)|M\right> \] \be \hspace{1cm}\times p(\a|M)
\prod_{i=1}^{3}\left\{ U_{\rm g}(M,\br_i-\bx,\a,\avec)\right\}\
;\ee
\[\zeta^{\rm 2H}_{\rm g}=
\frac{1}{\nbarg^3(8\pi^2)^2}\int \prod_{i=\{1,2\}} \left\{ dM_i\;
d \bx_i\; d{\bf a}_i\; d\avec_i\;n(M_i) p({\bf a}_i|M_i)\right.\]
\[ \hspace{1cm}\times \left. U(\bx_i-\br_i,M_i,{\bf
a}_i,\avec_i)\right\} \left<N(N-1)|M_1\right>\left<N|M_2\right> \; \]
\be \hspace{1cm}\times U(\bx_1-\br_3,M_1,{\bf a}_1,\avec_1)\;
\xi_2^s(1,2)+\cyc\ ;\ee
\[ \zeta^{\rm 3H}_{\rm g} =\frac{1}{(\nbarg8\pi^2)^3}\int
\prod_{i=1}^{3}\left\{ dM_i\; d \bx_i\; d{\bf a}_i\;d\avec_i \;
\left<N|M_i\right> \, n(M_i)\right. \] \be \hspace{1cm}\times\,
\left. p({\bf a}_i|M_i)\, U({\bf x}_i-\br_i,M_i,{\bf
a}_i,\avec_i)\right\} \,\zeta^s(1,2,3)\ .\ee
%


\subsection{Power spectrum \& bispectrum}\label{ssec:galpowbi}

We next write down the power spectrum and the bispectrum. As for the
analysis of the dark matter bispectrum, we define the useful functions
$W^{\rm g}_i$, $W^{\rm g}_{ij}$ and $W^{\rm g}_{123}$ by applying the
transformations described in Section \ref{ssec:galhalomodel} to
equations (\ref{eq:w1}), (\ref{eq:w12}) and (\ref{eq:w123}). Hence,
the galaxy power spectrum can be written as:
\be P_{\rm g}=P^{\rm 1H}_{\rm g}+P^{\rm 2H}_{\rm g} \label{eq:powerGal} \ ; \ee
\be  P^{\rm 1H}_{\rm g} = \frac{1}{\nbarg^2}\int
dM\,\left<N(N-1)|M\right> n(M)\, W^{\rm g}_{12}(M) \label{eq:power1H} \ ;\ee
\be P^{\rm 2H}_{\rm g} = \frac{P_{\L}}{\nbarg^2}
\prod_{i=1}^{2}\int dM_i \, \left<N|M_i\right>
n(M_i)\,b(M_i)\, W^{\rm g}_{i}(M_i)
\label{eq:power2H}\ .\ee
Similarly, the galaxy bispectrum can be written as:
\be B^{\rm g}_{123}= B^{\rm g1H}_{123}+B^{\rm g2H}_{123}+
B^{\rm g2H}_{123} \label{eq:biGal} \ ;\ee
\[ B^{\rm g1H}_{123}=\frac{1}{\nbarg^3} \int dM\,n(M)\left<N(N-1)(N-2)|M\right>\]
\be \hspace{1cm}\times\  W^{\rm g}_{123}(M)\
;\label{eq:bi1HGal}\ee
\[ B^{\rm g2H}_{123} = \frac{P_{\L}(k_2)}{\nbarg^3} \int\!\!\!
\prod_{i=\{1,2\}}\!\!\! \left\{dM_i\, n(M_i)\,b(M_i)\left<N(N-1)|M_1\right>
\right\}\]
\be \hspace{1cm}\times\, \left<N|M_2\right>\,
W^{\rm g}_{13}(M_1) \, W^{\rm g}_{2}(M_2) + \cyc\ ;\label{eq:bi2HGal}\ee
\[ B^{\rm g3H}_{123}=\frac{1}{\nbarg^3}\int \prod_{i=1}^{3}
\left\{dM_i \,\left<N|M_i\right>\, n(M_i)\,
W^{\rm g}_i(M_i)\right\} \]
\be \hspace{1cm}\times\ B^s_{123}(M_1,M_2,M_3) \label{eq:bi3HGal} \ .\ee
Owing to the fact that we are now modelling a discrete-point
distribution, as opposed to a continuous field, we must apply
corrections to both the power spectrum and bispectrum for shot noise.
For this we follow SSHJ and use relations that have the same form as
the Poisson case \citep{Peebles1980}:
\be P^{\rm g}_{c}(k)=P^{\rm g}(k)-\eta\ ;\ee
\be B_{c123}^{g}=B^g_{123}-\eta[P^g_{c1}+P^g_{c2}+P^g_{c3}]-\eta^2\ ;\ee
but where $\eta\equiv P^g(k\rightarrow\infty)$ instead of the usual
$\eta=1/\nbarg$.


\subsection{Central galaxy contribution}\label{ssec:central}

The above equations assume that the spatial distribution of the
galaxies follows the underlying dark matter profile. We may also wish
to impose the further constraint that, for every halo that contains at
least one galaxy, a single galaxy lies at the halo centre of mass.  In
terms of the halo model this means that there will be a small
modification to the clustering for terms that include the correlation
of objects within the same halo. We now derive how this affects the
small scale clustering.

Consider a triaxial halo of given mass $M$, with axis ratios $\a$,
orientation $\avec$ and with $N$ galaxies. For this halo we can
identify one central object and $N-1$ satellites.  For the two-point
statistics, there will be $(N-1)(N-2)/2$ pairs of satellite galaxies
and $N-1$ pairs that include the central galaxy and satellites. The
two-point clustering of the satellite galaxies can be calculated as per
usual. However, for the pairs that include the central galaxy, the
clustering will be weighted by the density profile. This owes to the
fact that the probability of finding a galaxy at position vector $\br$
from the central object now simply follows the density profile of
galaxies. Hence, for the 1-Halo term in the power spectrum, we have
the weight factor
\[ \CW_{12}^{2\rm g}(M,\a,\avec)= 2! \sum_{N>1}P(N|M)\left\{(N-1)
U_g(\bk_1,M,\a,\avec) \right.\] \be \hspace{0.5cm}\left.+
\frac{1}{2}(N-1)(N-2) U_g(\bk_1,M,\a,\avec)U_g^*(\bk_2,M,\a,\avec)
\right\}\ .\label{eq:satW12}\ee
Note that the factor of $2!$ in the above expression arises since the
term $U_gU_g^*$ is normalized by the total numbers of pairs with
double counting.  The galaxy window function is then obtained by
averaging the weight factor $\CW$ over the probability distribution
for the axis ratios and the Euler angles: e.g. in general the
$n$-point window function is given by
\be W^g_{12\dots n}(M)=\frac{1}{8\pi^2}\int d\a\, d\avec\, p(\a|M)
\CW^g_{12\dots n}(M,\a,\avec)\ .\label{eq:windowaverage}\ee
On computing the averages over $P(N|M)$ in equation (\ref{eq:satW12}),
we obtain
\[ \CW^{\rm 2g}_{12}(M,\a,\avec)=\left<N(N-1)|M\right>|U_g(\bk)|^2\]
\be \hspace{1cm}+2\,\left[\left<N-1|M\right>+P(0|M)\right]
\left\{U_g(\bk)-|U_g(\bk)|^2 \right\} \label{eq:wg12centsat} ,\ee
where we have suppressed the dependence of $U_{\rm g}$ on $\a$ and
$\avec$. The term $P(0|M)$, which is the probability of finding no
galaxies in the halo, arises because the sum over galaxy numbers
in equation (\ref{eq:satW12}) does not start from zero. Hence we
have terms like
\be \sum_{N>1}(N-1)P(N|M)=\left<N-1|M\right>-P(0|M)\ .\ee
On combining equations (\ref{eq:powerGal})-(\ref{eq:power2H}),
(\ref{eq:windowaverage}) and (\ref{eq:wg12centsat}), we arrive at a
complete description of the galaxy power spectrum.

Similarly, for the bispectrum we will have $(N-1)(N-2)/2$ triplets
that include the central galaxy and $(N-1)(N-2)(N-3)/3!$ triplets that
comprise only satellite galaxies. The 2-Halo term in the bispectrum
will be modified exactly as described for the power spectrum. However,
for the 1-Halo term the weight factor must be modified as follows:
\[ \CW^{\rm 3g}_{123}(M,\a,\avec)= 3! \sum_{N>2}P(N|M)
\left\{\frac{1}{2}(N-1)(N-2)
\prod_{i=1}^{2}U_{\rm g}(\bk_i)
\right.
\]
\be \hspace{1cm}\left. + \frac{1}{6}(N-1)(N-2)(N-3)
\prod_{i=1}^{3}U_{\rm g}(\bk_i)\ \right\}\ ,\label{eq:3pointCentWeight}\ee
where we have included the factor of $3!$ to account for the fact that
the profiles are normalized by the total triplets, which are counted
in a way that does not prevent repetition. On computing the average
over $P(N|M)$ we get
\[\CW^{\rm 3g}_{123}(M,\a,\avec)=
\left<N(N-1)(N-2)|M\right> \prod_{i=1}^{3}U_{\rm g}(\bk_i) \]
\[\hspace{1cm}+\,3!
\left\{\frac{1}{2}\left<N(N-1)\right>-\left<N-1|M\right>-P(0|M)
\right\}\]
\be \hspace{1cm}\times\ \left[\prod_{i=1}^{2}U_{\rm g}(\bk_i)-
\prod_{i=1}^{3}U_{\rm g}(\bk_i)
\right]\label{eq:wg123centsat}\ .\ee
Thus on combining equations (\ref{eq:biGal})--(\ref{eq:bi3HGal}),
(\ref{eq:windowaverage}) and (\ref{eq:wg123centsat}) we also arrive at
a complete analytic description of the galaxy bispectrum with one
central galaxy and $N-1$ satellites.

%
\begin{figure}
\centerline{
\includegraphics[width=8cm]{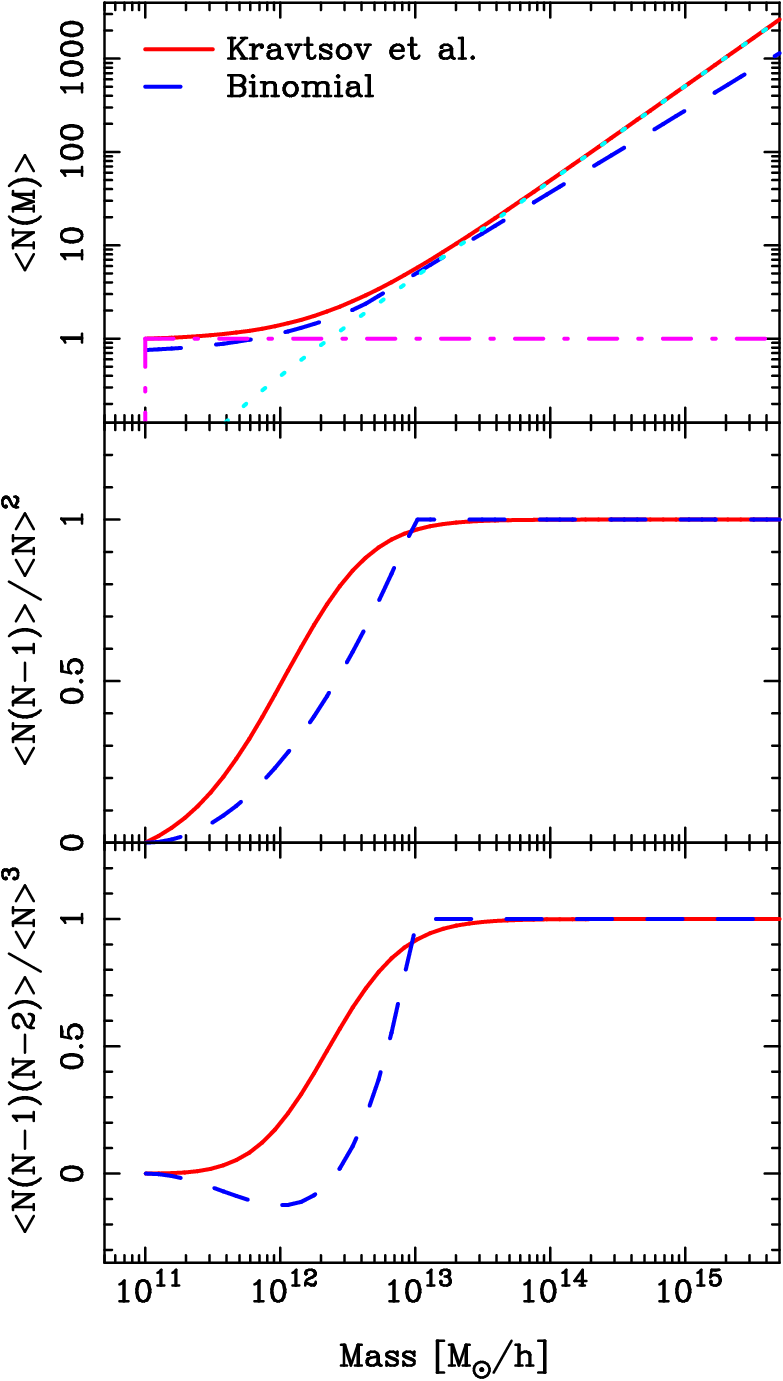}}
\caption{\small{Comparison of the first three moments of the halo
occupation probability function of K04, solid lines, with those of
SSHJ, dashed lines. In the top panel, the
dot-dash line represents the central galaxy contribution and the dotted
line represents the satellite galaxy contribution.}
\label{fig:BinMoments}}
\end{figure}
%

\subsection{Halo Occupation Distribution}\label{sec:galmodel}

To calculate the galaxy bispectrum we require a specific model for the
mean spatial distribution of galaxies within each halo and also the
first three moments of the halo occupation probability
function. Firstly, we will make the usual assumption that the galaxy
density distribution follows the mass profile, and hence
$U_g(\br,M)\propto U(\br,M)$. In accordance with the ideas of the
previous section, we will consider the effects on the bispectrum
caused by placing one galaxy at the centre of each halo.  Secondly,
for $P(N|M)$, we will consider two potentially equally viable models
from the literature: the first is the binomial model proposed by SSHJ,
that was constrained to match the predictions of the semi-analytic
models of \citet{Kauffmannetal1999}; the second is the model developed
by \citet[][hereafter K04]{Kravtsovetal2004}. In this model it is
assumed that if the halo mass lies above some minimum mass threshold,
then there is always one central galaxy. The remaining $N-1$ galaxies
are then taken to be satellites of the central galaxy, and it was
found that these follow a Poisson process \footnote{Note that the term
``central galaxy'' used in K04 is somewhat unfortunate given our
discussion in the last section, since the occupation probability model
of K04 in fact makes no prediction for the spatial distribution of
galaxies within the halo. Thus this ``central galaxy'' may, in
principle, lie at any point within the halo boundary.}. The full
details of both models are summarized in Appendix \ref{app:pN}.

In Figure \ref{fig:BinMoments} we show the differences between the
first three factorial moments of the two probability functions.  Note
that in the limit of high masses, the power-law index of the first
moment for the model of K04 is $\left<N\right>\propto M^{\alpha}$,
with $\alpha\sim1.0$, whereas for the binomial model it is
$\alpha\sim0.9$. In this limit the second and third factorial moments
scale identically with the mean $\left<N\right>$ for both models. For
halo masses below $\sim 10^{13}$ M$_{\sun}$ (i.e. group and galaxy
masses) the second and third moments are sub-Poisson in both models.
Furthermore, for a given mass the binomial $P(N|M)$ is much narrower
than for the K04 model.


\subsection{Results: Galaxy Bispectrum}


\begin{figure*}
\centerline{ 
\includegraphics[width=7.cm]{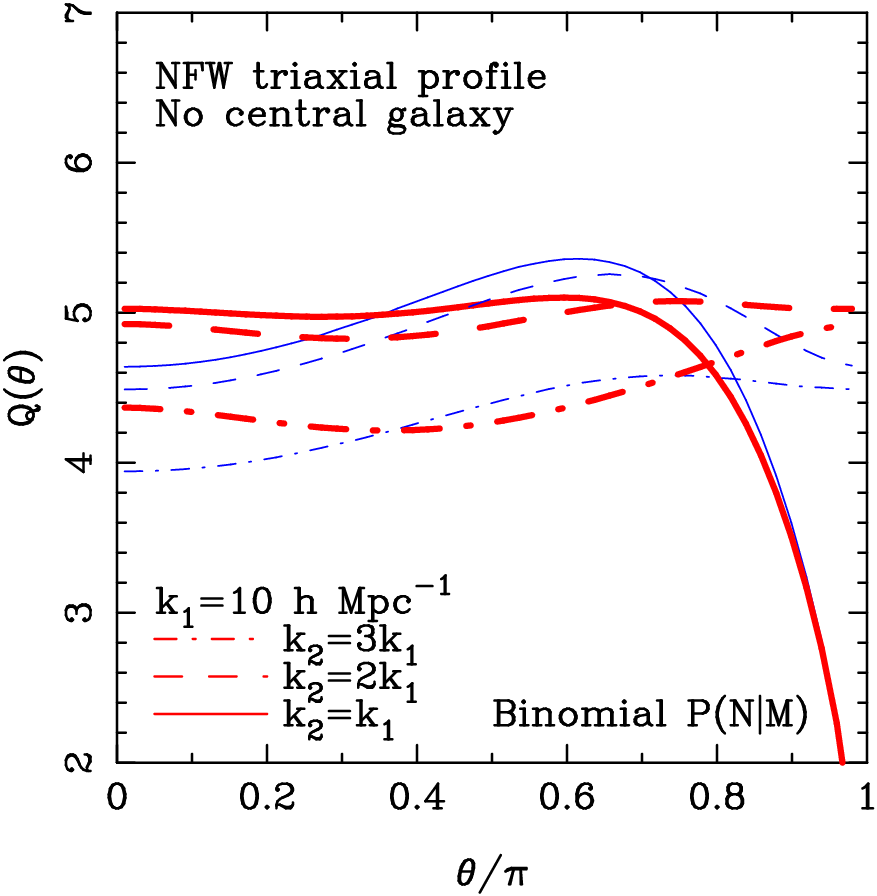}
\includegraphics[width=7.cm]{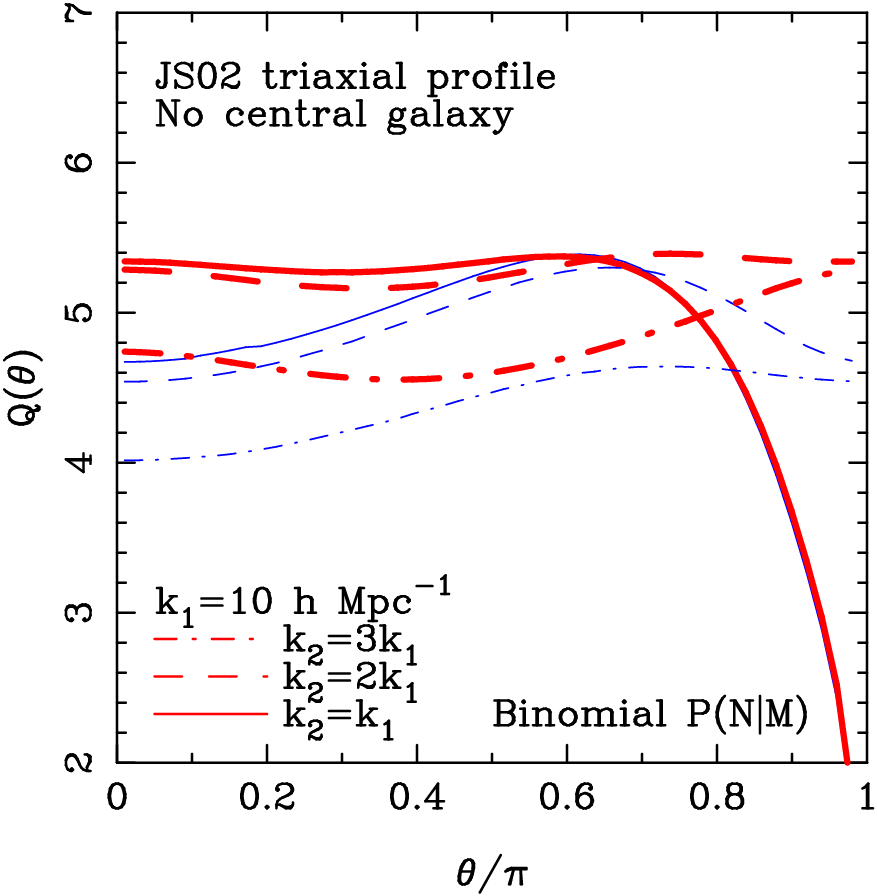}} 
\centerline{
\includegraphics[width=7.cm]{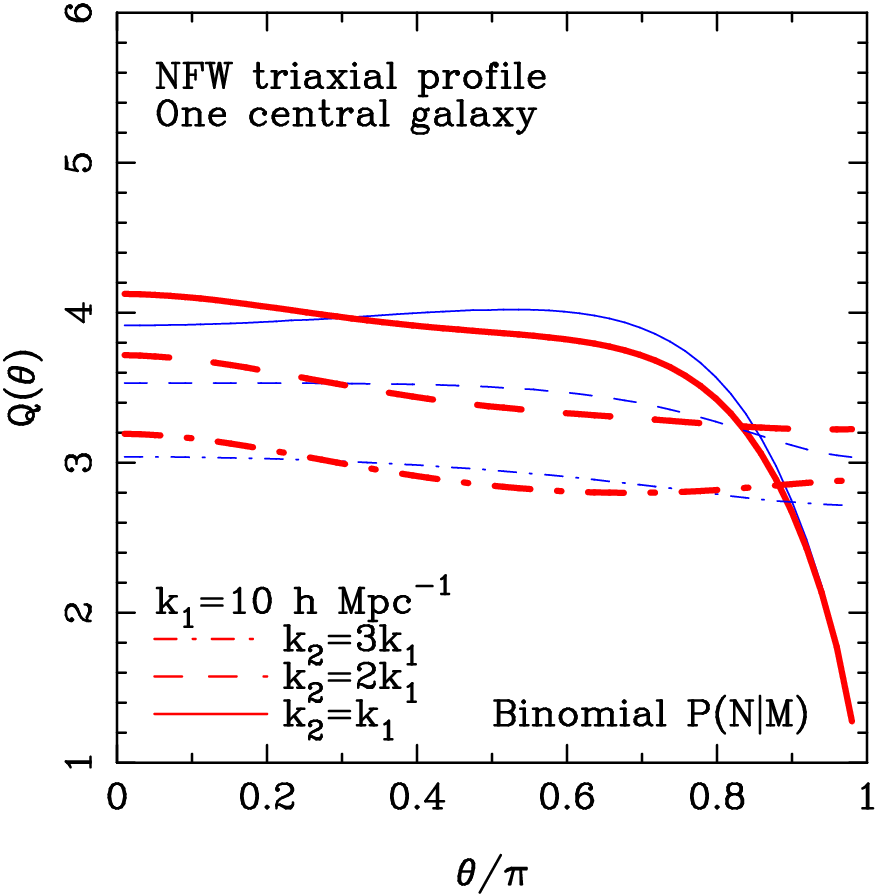}
\includegraphics[width=7.cm]{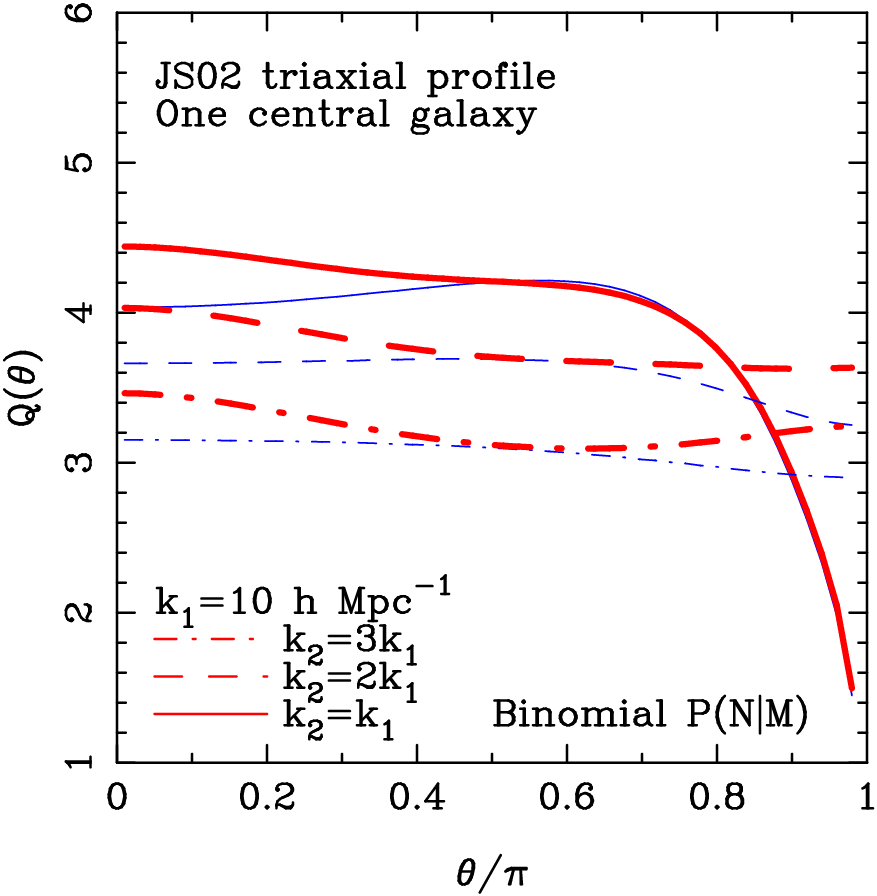}}
\caption{\small{Configuration dependence of the galaxy bispectrum. The
halo occupation probability function is taken to be the binomial model
of SSHJ. Line styles are the same as in Figure \ref{fig:Qtheta2}. }
\label{fig:galbi1}}
\end{figure*}


\begin{figure*}
\centerline{ 
\includegraphics[width=7.cm]{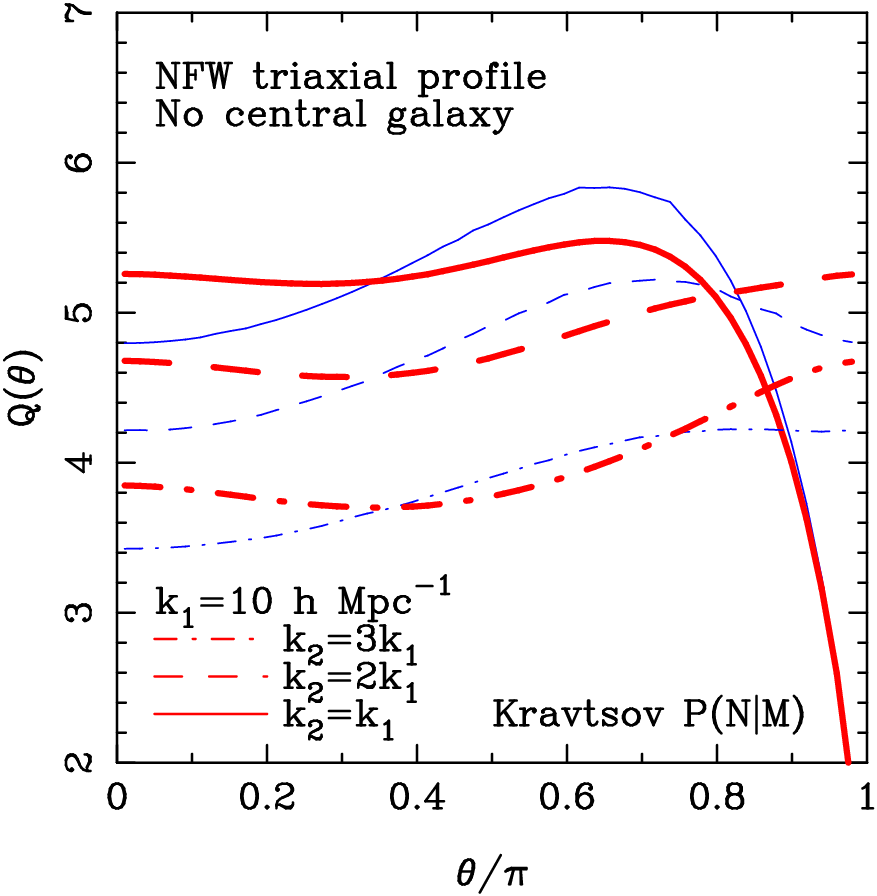}
\includegraphics[width=7.cm]{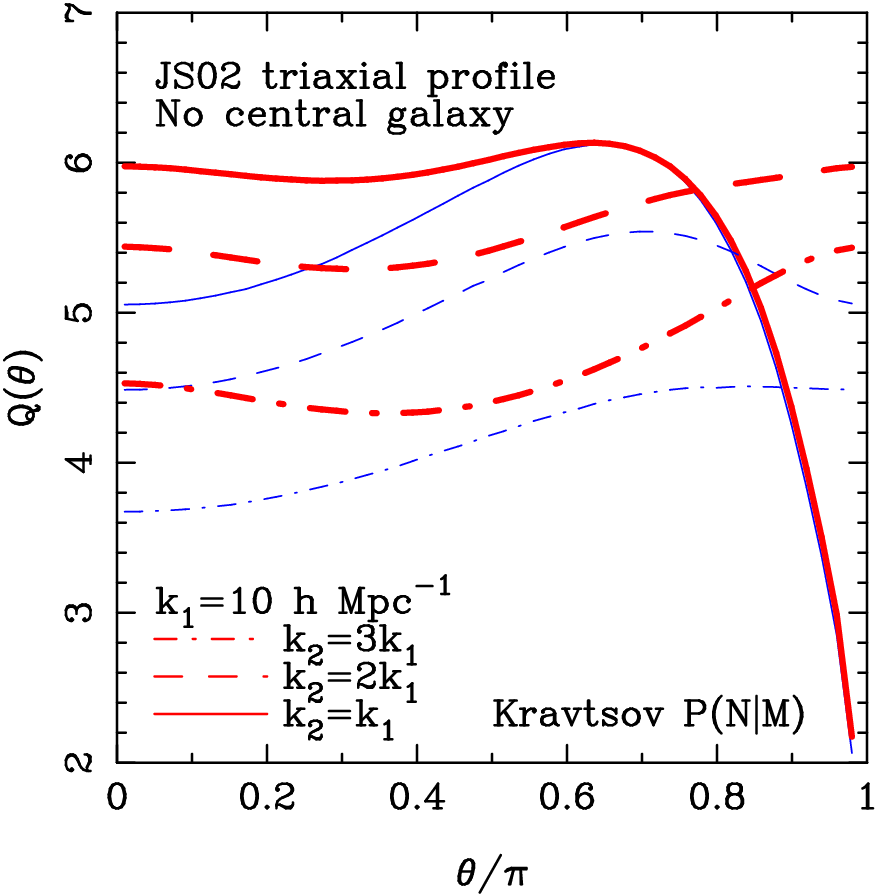}} 
\centerline{
\includegraphics[width=7.cm]{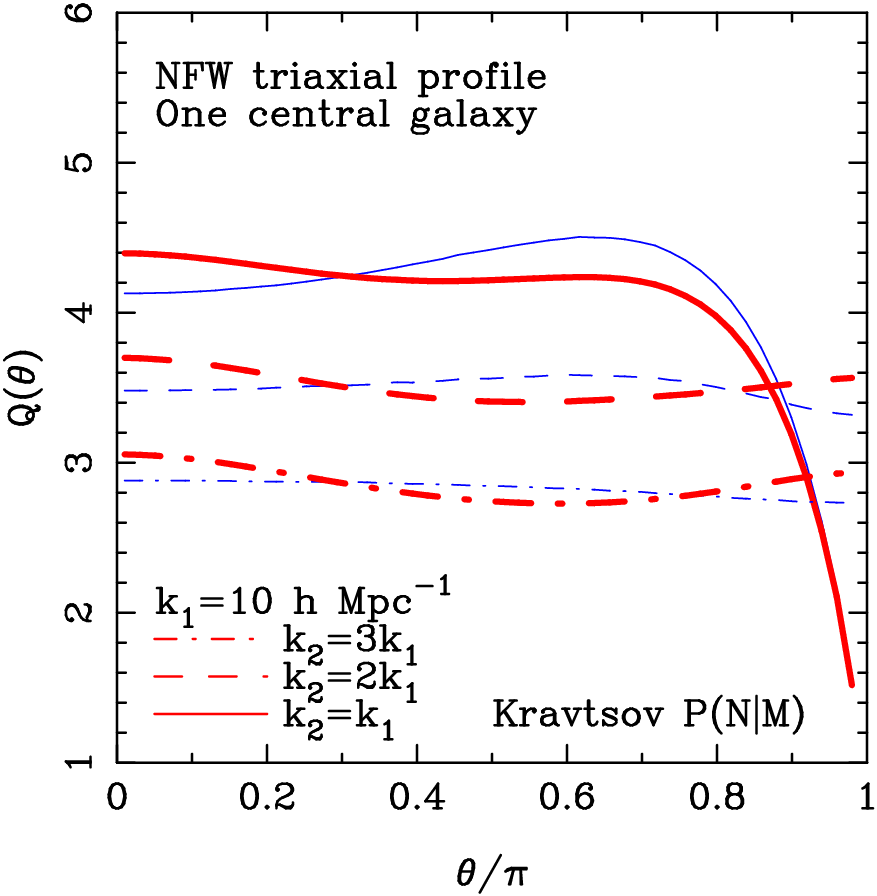}
\includegraphics[width=7.cm]{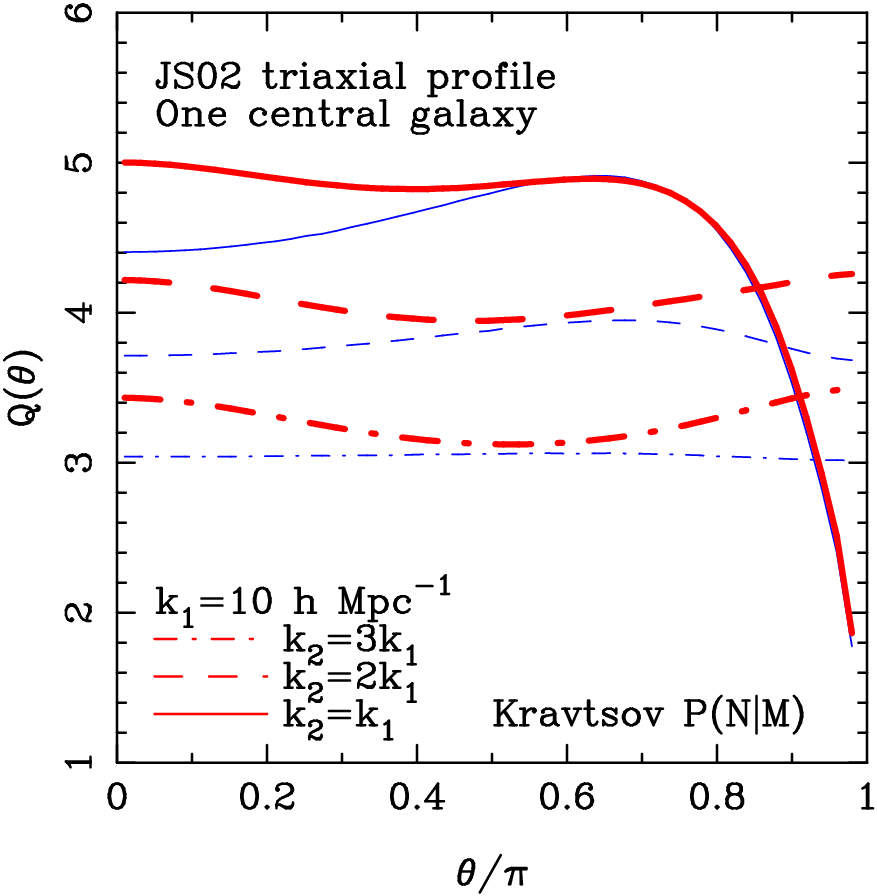}}
\caption{\small{Same as Figure \ref{fig:galbi1}, however, this time
the halo occupation probability function is taken to be the Poisson
satellite model of K04. Again, line styles are the same as in Figure
\ref{fig:Qtheta2}.}
\label{fig:galbi2}}
\end{figure*}


Figures \ref{fig:galbi1} and \ref{fig:galbi2} show the triaxial and
spherical halo model predictions for $Q(\theta)$ for the binomial
model of SSHJ and the Poisson satellite model of K04, respectively. In
generating these predictions we chose the minimum halo mass for the
binomial model to be $M_{\rm min}=10^{11} h^{-1}\ M_{\odot}$, which
for the assumed cosmology gave a number density of galaxies
$\nbarg=3.81475\times 10^{-2} h^3 {\rm Mpc}^{-3}$. In order to compare
the two occupation probability models consistently, it was important
to remove any scaling due to the overall numbers of galaxies, since
$Q\propto \nbarg$. This was done by finding the $M_{\rm min}$ in the
K04 model that reproduced the same galaxy number density as for
SSHJ. We found this to be $M_{\rm min}=1.28\times 10^{11} h^{-1}\
M_{\odot}$.

The top panels of both figures show the predictions for the case where
all galaxies are distributed within the haloes according to the
density profile (Section \ref{ssec:galpowbi}). The bottom panels show
how the predictions change when one galaxy is assumed to be located at
the centre of each halo (Section \ref{ssec:central}).  As in Section
\ref{sec:results} we show results for both the triaxial NFW profile
(left panels) and the JS02 model (right panels). Once again, it is
seen that the triaxial halo model predictions for $Q(\theta)$ differ
quite substantially from those of the spherical model. The
characteristic features of these predictions are essentially the same
for all of the eight panels in the figures: the triaxial model
produces a boosted signal for colinear triangle configurations and a
suppression for configurations that are close to equilateral. However,
and importantly, the functional form of these curves appears not to be
strongly dependent on the precise form of the chosen HOD.

On comparison of these results with the mass bispectrum
(Fig. \ref{fig:Qtheta2}), we see that the overall effect of
triaxiality is weaker in the galaxy statistics. For the binomial model
the effects are reduced by $\sim50\%$, and for the model of K04 they
are reduced by roughly $\sim30\%$. The better sensitivity of the K04
model can be attributed to the fact that, for high mass haloes, the
power-law index of the first moment of $P(N|M)$ is a steeper function
of mass than the binomial model: as mentioned previously, $\alpha \sim
1 $ for the K04 model and $\sim 0.9$ for the binomial model. In other
words the most massive haloes, which are also the most triaxial, will
host relatively more galaxies in the K04 model than in the binomial
model.

Comparison of the upper and lower panels of Figures \ref{fig:galbi1}
and \ref{fig:galbi2} reveals the overall importance of having one
central galaxy. The first point to note is that, for both $P(N|M)$
models and density profiles considered, the overall effect is that
$Q(\theta)$ is shifted by a factor $\sim1$.  After some tests, we
found that this behaviour was a consequence of the bispectrum being
normalized by products of the power spectrum (equation \ref{eq:Q}). To
see this consider the following: on small scales, the power spectrum
is more strongly weighted to contributions from low mass, low
occupancy halos; this is due to the $\left<N(N-1)|M\right>$ weighting
in the 1-Halo term. Whilst both the power spectrum and the bispectrum
are sharply enhanced by the presence of the central galaxy, the power
spectrum is relatively more enhanced owing to the added numbers of
lower mass haloes that contribute to the signal; these are not
included in the bispectrum as only haloes that host at least three
galaxies can contribute. The result then follows.

Secondly, the central galaxy also appears to reduce the overall affect
of halo triaxiality on the clustering. This can be understood by
considering haloes that host three galaxies: when one is placed at the
centre we find that the 1-Halo term is $B(k)\propto
\left[U_g(k)\right]^2$ (equation \ref{eq:3pointCentWeight}), whereas
for the case of no central galaxy, $B(k)\propto
\left[U_g(k)\right]^3$. As was shown in Figure \ref{fig:TriaxWindow},
the window functions become more sensitive to the halo shape as more
points are placed in the $k$-space halo.  Thus, placing one galaxy at
the halo centre effectively removes one of these points.

As a final remark, we note that the combination of measurements for the
triangle configurations with $k_2/k_1=1$, 2 and 3, in both the
spherical and triaxial models, appear to depend strongly on the
HOD. In particular, the spread in the amplitude of $Q$ as the ratio
$k_2/k_1$ changes is substantially greater for the model of KS04,
while the predictions for the SSHJ model appear closer to those of the
mass bispectrum (cf Figure \ref{fig:Qtheta2}). The response of these
predictions to the presence, or otherwise, of a central galaxy is also
very different in the two models, with the SSHJ model being affected
to a much greater extent by this assumption.

In summary, on small scales the reduced galaxy bispectrum has been
shown to be a sensitive function of the halo shape and density
structure, the occupation probability $P(N|M)$, and the central galaxy
assumption. We note that the functional form of $Q(\theta)$ is most
strongly affected by the shapes of the haloes themselves and is not
greatly affected by the exact form of the HOD. However, the amplitude
of each prediction, relative amplitudes of predictions made over
different scales and the strength of the triaxiality effect are all
sensitive to the particular choice of HOD.


\section{Conclusions}\label{sec:conclusions}

In this paper we have explored how sensitive the real space bispectrum
in cosmology is to the underlying shapes of dark matter haloes, for
both matter fluctuations and galaxies. We achieved this through
application of the triaxial halo model of SW05, adapting it to the
problem of higher order statistics in a straightforward manner. In our
model we accounted for dark matter haloes that have some known
distribution of shapes and that are randomly orientated in space.

Analytic expressions were written down for the real space three-point
matter correlation function and bispectrum. We then concentrated on
the bispectrum and found that the 1-, 2- and 3-Halo terms could be
compactly expressed as 7-D integral equations. These were solved
numerically using an efficient multi-dimensional quadrature routine
\citep{Korobov1963,Conroy1967}. Two models for the density profile of
the triaxial haloes were considered: the first was the triaxial NFW
model of SW05, which allowed us to explore the affects of halo shape
alone on the bispectrum; the second was the more realistic model of
JS02. The bispectrum was examined over a wide range of scales and for
different configurations of $k$-space triangles.

We found that for equilateral triangle configurations the effect of
halo triaxiality on the bispectrum was to produce a suppression on
scales $k>0.2\, h \,{\rm Mpc}^{-1}$, relative to the spherical halo
model predictions. This suppression was at the level of $<7\%$ for
both density profiles.  The reduced bispectrum was considered next, as
function of $k$-space triangle configuration. We found that on large
scales this quantity was insensitive to halo shapes. However, on small
scales the predictions for the spherical and triaxial halo models
differed significantly in both the amplitude and functional form of
$Q(\theta)$. In general we found that, relative to the spherical
model, the triaxial halo model predictions produced excess signal for
triangle configurations that were colinear and non-vanishing, and a
deficit for triangles that were close to equilateral. We observed that
the overall form of $Q(\theta)$ depends little on the density profile
model itself, modulo an amplitude shift, but rather that it is the
shape of the haloes that dictates the shape of the curves.

Our results were then considered in relation to the discrepancies
between the $N$-body simulations and the predictions from the
spherical halo model that were noted by SSHJ and
\citet{Fosalbaetal2005}. We suggested that, for the case of the
equilateral triangle configurations, the inconsistencies between
theory and simulations are not due entirely to a break down in the
spherical approximation.  Rather, they are likely due to the
combination of finite volume effects in the simulations and also other
corrections to the halo model (SSHJ; Wang et al. 2004; Fosalba, Pan \&
Szapudi 2005). However, for the case of the configuration dependent
bispectrum it was shown that the discrepancies could be well
attributed to the break down of the spherical halo approximation,
modulo an amplitude offset, which arises from the correction required
to reconcile halo model predictions for the equilateral bispectrum
with numerical simulations.

In the second part of this paper, we explored how halo shapes affect
the real space galaxy bispectrum. Galaxies were included in the halo
model in the usual way (Seljak 2000; SSHJ; Berlind \& Weinberg 2002),
and we considered two different schemes for the halo occupation
probability function. The first of these was the binomial model
developed by SSHJ, while the second was the Poisson satellite model of
K04. For both probability functions, the predictions of the
configuration dependent, reduced bispectrum were found to be very
similar to those for the mass.  However, the overall magnitude of the
effect was reduced by $\sim50\%$ for the binomial model, and
$\sim30\%$ for the K04 model. It was argued that the K04 model was
more sensitive to halo shapes because of the stronger dependence on
halo mass for the first moment of the occupation probability function.
We also showed that the predictions for the binomial model possessed a
stronger dependence on the presence of a central galaxy, when compared
to the model of K04. Importantly, we found that the essential effects
of triaxiality on the reduced bispectrum did not depend strongly on
the exact details of the halo occupation distribution. However, a
combination of measurements on different scales are sensitive to the
both halo shape and the HOD.

We conclude that the bispectrum provides significant information about
the shapes of dark matter haloes in addition to the (galaxy and mass)
halo density profiles, and the halo occupation probability function.
Consequently, in order to use the halo model to make precise
constraints on the HOD, it will be necessary to take into careful
consideration halo triaxiality. We remark, however, that our study did
not account for the effect of redshift space distortions, which will
have a significant impact on the behaviour of the galaxy bispectrum,
particularly on small scales. Furthermore, we have also neglected the
changes to the bispectrum that are caused by selecting different galaxy
populations i.e., populations based on colour, luminosity or
type. Neither have we considered realistic selection functions, or
survey geometry.  We reserve such investigations for future work, when
we will study in more detail the bispectrum as a tool for measuring
the HOD.

It remains to be seen whether measurements of the three-point
statistics in weak lensing will be sensitive to halo shapes. An
initial study of this problem has been conducted by
\citet{HoWhite2004}, who calculated the three-point shear correlation
function for triaxial haloes and found strong effects.  The situation
they considered was rather idealized, however, and was not a full
halo-model calculation. Nevertheless, the potential for measuring such
effects is clear.  Recent work by \citet{TakadaJain2003b} showed, for
an 11 square degree mock cosmic shear survey, that the three-point
shear correlation function from ray-tracing simulations was in good
agreement with the predictions of the spherical halo model. However,
they noted that on small scales there were significant and unexplained
discrepancies. For forthcoming weak lensing surveys which aim to cover
even larger fractions of sky ($>100$ square degrees), the shear
three-point functions will be accurately determined over a wide range
of scales. Discrepancies between halo model predictions and
observation will become very significant and important to understand.
Moreover, the detection of shape information in the mass bispectrum or
correlation function via weak lensing would be an important
consistency test of the CDM paradigm, since it has been shown that
plausible variants of the dark matter model, like SIDM and WDM, will
give rise to a different spectrum of dark matter halo shapes
\citep{AvilaReeseetal2001,Yoshidaetal2000}.

Lastly, we mention that a recent study of the shapes of dark matter
haloes in numerical simulations with gas cooling by
\citet{Kazantzidisetal2004} has shown that the formation of central
condensations may act to sphericalize the inner regions of haloes.  At
the virial radius, where halo shapes are usually measured, such
processes will have relatively little effect. However, in order to
achieve precise predictions to compare with observation, it will be
necessary to quantify in detail how these effects modify halo profiles
and axis ratio distributions.


\section*{acknowledgements}

We thank Shirley Ho, Masahiro Takada and Roman Scoccimarro for useful
discussions and Peter Schneider for useful comments on an early draft
of the paper. RES acknowledges the kind hospitality of the University
of Bonn where part of this work took place. RES was supported by the
National Science Foundation under Grant No. 0520647. PIRW was
supported by the Deutsche Forschungsgemeinschaft under the project
SCHN 342/6--1, and within the DFG Priority Programm 1177 `Witnesses of
Cosmic History' by the project SCHN 342/8--1.


\setlength{\bibhang}{2.0em}


\appendix


\section{Rotation matrix}\label{app:rotation}
Owing to there being several equivalent ways to define the Euler
angles for the rotation matrix ${\mathcal R}(\alpha,\beta,\gamma)$, we
make explicit the definition that we use throughout. The matrix for
the $z-y'-z''$ rotation is given by:

\be {\mathcal R}(\alpha,\beta,\gamma)\equiv \left(
\begin{array}{ccc}
\left( C_{\beta}\, C_{\alpha} C_{\gamma}\right. & \left( C_{\beta}\, S_{\alpha}\,
C_{\gamma} \right. &
-S_{\beta}\, C_{\gamma}\\
\left.-S_{\alpha}\,S_{\gamma}\right) & \left.+C_{\alpha}\,S_{\gamma}\right)& \\
\left(-C_{\beta}\, C_{\alpha}\, S_{\gamma} \right.& \left(-C_{\beta}\, S_{\alpha}
\,S_{\gamma} \right. & S_{\beta}\, S_{\gamma} \\
\left.-S_{\alpha}\,C_{\gamma}\right) & \left.+C_{\alpha}\,C_{\gamma}\right)& \\
S_{\beta}\, C_{\alpha} & S_{\beta}\, S_{\alpha} & C_{\beta} \\
\end{array}\right)
\ ,\ee
where we have adopted the short hand notation $C_{x}=\cos
x$ and $S_{x}=\sin x$.


\section{Occupation probability}\label{app:pN}

\subsection{Scoccimarro et al: binomial distribution}

We now summarize the details of the occupation probability model
proposed by SSHJ to describe the results from the semi-analytic galaxy
formation models of
\citet{Kauffmannetal1999}. \citet{ShethDiaferio2001} measured the
first two moments of $P(N|M)$ from these models.  For the first moment
of they found
\be \left<N|M\right>=\left<N_B|M\right>+\left<N_R|M\right>:\ee
\be
\left<N_B|M\right>=0.7\left(\frac{M}{M_B}\right)^{\alpha_B}\ ;
\hspace{0.5cm}
\left<N_R|M\right>=\left(\frac{M}{M_R}\right)^{\alpha_R}\ ,\ee
where $N_B$ and $N_R$ represent the numbers of red and blue galaxies.
The blue galaxy parameters are $\alpha_B=0$ for haloes with masses in
the range $10^{11}M_{\odot} h^{-1}<M<M_B$ and $\alpha_B=0.8$ for
$M>M_B$, where $M_B=4.0\times10^{12}M_{\odot} h^{-1}$.  The red galaxy
parameters are $\alpha_R=0.9$ and $M_R=2.5\times10^{12}M_{\odot}
h^{-1}$ for $M>10^{11} h^{-1}M_{\odot}$.
For the second moment, they found that a sub-Poissonian distribution
was preferred for low-mass haloes and an almost Poisson distribution
for higher-mass haloes. They characterized this through
\be \left<N(N-1)|M\right>=\lambda^2(M)\left<N|M\right>^2, \ee
where the function $\lambda$ expresses the deviation from a Poisson
probability distribution and has the form
\be \lambda(M)\approx\left\{
\begin{array}{ll}
\log_{10}(M/M_{11})^{1/2}\ & M\le10^{13}M_{\odot}\,h^{-1}\\
1\ &M>10^{13}M_{\odot}\,h^{-1}
\end{array}\right.
\ee
where $M_{11}\equiv10^{11}M_{\odot}\,h^{-1}$.

SSHJ then proposed that, since $P(N|M)$ was not well described by a
Poisson process, a better description might be afforded through the
binomial distribution:
\be P(N=n|M)={\mathcal C}_n^{\Nmax}\,p_M^{n}(1-p_M)^{\Nmax-n}\ ,\ee
where ${\mathcal C}_n^{\Nmax}=\Nmax!/(\Nmax-n)!\,n!$. The binomial
probability function is completely specified by two free parameters:
the probability that out of one trial the event occurs $p_M\equiv
p(M)$; and the total number of trials $\Nmax\equiv N^{\rm max}(M)$.
These parameters were then fixed by matching the first and second
moments to those measured from the semi-analytic galaxies. This gives
\be p_M=\left<N|M\right>\left[1-\lambda^2(M)\right] \ , \hspace{0.5cm}
\Nmax=\left[1-\lambda^2(m)\right]^{-1}\label{eq:binomialpar}.\ee

Having specified $P(N|M)$ one may then calculate the galaxy clustering
to any desired order. Of particular interest are the factorial
moments. For the binomial distribution these are most readily obtained
by differentiation of its frequency generating function:
\[
\left<N(N-1)\dots(N-j)|M\right> \equiv
\left.\frac{d^jF(t)}{dt^j}\right|_{t=0}\] \be \hspace{2cm} =
\left.\frac{d^j}{dt^j}\left\{p_M t+1-p_M\right\}^{\Nmax}\right|_{t=0}
\ .\ee
On performing the differentiation and substituting in for the
relations (\ref{eq:binomialpar}), one finds the general relation
(SSHJ)
\[\left<N(N-1)\dots(N-j)|M\right>=\]
\be
\hspace{2cm}\lambda^2(2\lambda^2-1)\dots(j\lambda^2-j+1)\left<N|M\right>^{j+1}\
\label{eq:factmom}.\ee
As a final remark we note that this particular occupation function is
somewhat problematic to implement in practice: in order to sample from
it, one requires that $\Nmax$ be an integer. In an analytic
calculation this is not the case since one obtains the factorial
moments directly from equation (\ref{eq:factmom}).  Indeed, as
$\lambda$ is a continuous variable, so to will be $\Nmax$.  Thus when
$\Nmax$ is small there are significant differences between the sampled
and analytic distributions.


\subsection{Kravtsov et al: central and satellite split}

Much recent attention has focused on the separation of central and
satellite galaxies, and characterizing the statistical properties of
these objects as distinct populations (K04). The occupation
probability under this prescription is therefore re-written
\be P(N|M)=\sum_{l=0}^{N}P_{\cent}(l|M)P_{\sat}(N-l|M)\label{eq:prob1}\ee
where $P_{\cent}(l|M)$ and $P_{\rm sat}(N-l|M)$ are the probabilities
for getting $l$ central galaxies and $N-l$ satellite galaxies,
respectively.

In the work of K04 it was supposed that there was a one-to-one mapping
between haloes (parent or substructure) and galaxies (central or
satellite), and also that there was always only one central galaxy
associated with the parent halo. This then lead to the following
statement:
\be P_{\cent}(N=1|M)=
\left\{
\begin{array}{ll}
1 & \hspace{1cm} M\ge M_{\rm min} \\
0 & \hspace{1cm} M<M_{\rm min}.
\end{array}
\right.
\ee
Under this condition equation (\ref{eq:prob1}) simply becomes
\be P(N+1|M)=P_{\sat}(N) \ ;\hspace{1cm} M\ge M_{\rm min}\ .\ee
Following this one may write down how the moments of the full halo
occupation probability $P(N|M)$ are related to the moments of the
satellite population $P_{\sat}(N|M)$.

K04 went on to show that the first moment of the satellite galaxy
probability function was well characterized as a single power-law,
$\left<N\right>_{\sat}\propto M$, with a steeper decline for lower
halo masses.  The exact function they found was
\be \left<N\right>_{\sat}=\left(\frac{M}{M_1}-C\right)^{\beta}\ , \ee
where $\beta=1.03$, $C\approx0.045$ or $M_1/M_{\rm min}=22$, where
$M_{\rm min}$ is the minimum halo mass. Furthermore, it was found that
the second and third moments of the satellite galaxy distribution were
well characterized by a Poisson process. Thus, the first three
factorial moments of the full halo occupation probability function
are:
\be \left<N\right>=\left<N_s\right>+1\ ;\ee
\be \left<N(N-1)\right>=\left<N_s\right>^2-1\ ;\ee
\be \left<N(N-1)(N-2)\right>=\left<N_s\right>^3-3\left<N_s\right>+2\ .\ee

\end{document}